\documentclass[11pt]{article}

\usepackage[a4paper,margin=1in]{geometry}
\usepackage{amsmath,amssymb,array,booktabs,flafter,float,graphicx,longtable,microtype,xcolor}
\usepackage{placeins}
\usepackage{tikz}
\definecolor{legbg}{RGB}{23,36,52}
\usepackage[colorlinks=true,citecolor=blue,linkcolor=black,urlcolor=black]{hyperref}
\hypersetup{
  pdftitle={Percolation of Data Constraints and Reorganization of Boolean Function States in Parity Learning},
  pdfauthor={Han Ma}
}

\newcommand{\Tr}{\operatorname{Tr}}
\newcommand{\Kf}{K_f}
\newcommand{\wstar}{w_\star}

\newcommand{\Ablk}{A_{\rm blk}}

\newcommand{\Gammone}{\Gamma_1}
\newcommand{\dfree}{d_{\rm free}}

\title{Percolation of Data Constraints and Reorganization\\
of Boolean Function States in Parity Learning}
\author{Han Ma\\
\small CPHT, CNRS, École Polytechnique, Institut Polytechnique de Paris,\\
\small Palaiseau 91128, France}
\date{}

\begin{document}
\maketitle

\begin{abstract}
We study how a neural network reorganizes its learned function as training proceeds and as
previously unseen parity data accumulate. The trained model is itself a Boolean function on binary
strings of finite length. We then evaluate it on all data, inside and
outside the training set, and read off its complete sign table. 
Its $\pm1$ entries are the
amplitudes of a normalized state, dubbed the function-state, whose inverse
subsystem purity $\Kf=1/\Tr(\rho_A^2)$ is an effective Schmidt rank. We find that small data additions 
broaden the Schmidt spectrum, whereas larger additions
reconcentrate it in the target mode. The former is called differentiation because the
model distinguishes inputs more sharply. The latter is called
target-selective recoherence, where the model settles on the single rule shared by the
added data.
As the dataset grows, the connectivity of the constraint graph defined by the training set increases
monotonically, whereas the learned inverse purity changes nonmonotonically.
On a fixed connected dataset, as learning proceeds, the Schmidt spectrum
begins to reorganize, followed by gains in target alignment and later by improvement in predictive
accuracy. This ordering distinguishes three aspects of learning:
constraint identifiability, function-state organization, and predictive accuracy.
At fixed data size, a function that has already been learned can
re-differentiate, an event we call retreat, without changing graph connectivity. Recovery reconcentrates the Schmidt spectrum in a
similar target-selective pattern, without exact recurrence of the Boolean state.
Thus the percolation of data constraints fixes the identifiable support,
while the reduced state produced by training continues to reorganize even after that support stops changing.

\end{abstract}

\section{Introduction}
\label{sec:introduction}

In statistical physics, computational complexity, and quantum
information, many problems ask how accumulating local constraints reorganize a global space
of solutions or states. Classical satisfiability problems make two structures
explicit: clauses form a constraint graph,
while the assignments satisfying every clause form a solution space, whose
geometry is set by the overlap of any two assignments and by their clustering. As clause density increases, graph connectivity and 
the size of the giant component can change at values distinct from those of clustering, freezing, and the
eventual loss of satisfiability
\cite{mezard2009information}. Quantum satisfiability lifts the same two-structure split into
Hilbert space: the support hypergraph records which qubits each local
term acts on, while the satisfying subspace
consists of states compatible with every term. Increasing clause density can
alter the dimension, product structure, and entanglement of this subspace
before satisfiability is lost
\cite{laumann2012statistical}. Accordingly, the structural changes in the underlying constraint 
network is distinct from the changes in the states it supports.

Supervised learning presents a related constraint-selection problem.
In the realizable setting studied here, a fixed target function generates every label. Each labeled
data point imposes a constraint. Any function consistent with the training data must assign
the observed label to that input. Accumulating data therefore contracts the compatible function space, 
which remains nonempty because it always contains the target function.
The remaining freedom depends on both the observed data and the class of functions
the model can represent. Fitting the training labels alone does not establish that
the target function has been learned. It fixes the outputs on the training
inputs but does not, by itself, determine predictions on unobserved inputs.
Different training runs can achieve similar accuracy while making different predictions
on unobserved inputs~\cite{geiger2020scaling}. Accuracy therefore does not fully
characterize the learned function. The ensemble of such learned functions over independent training runs is studied with
classical replica overlaps in a companion paper~\cite{ma2026replica}. Biases toward simpler functions further motivate
studying which compatible function training selects~\cite{valleperez2019deep}.

After the observed data have been fit, we ask whether the learned function can
continue to reorganize and how its organization responds when previously unseen data are added. We find two forms
of dynamics. When previously unseen data points are added, the
learned function reorganizes nonmonotonically even as the compatible function space
contracts. On a fixed dataset, the function continues to
reorganize, moving away from and later back toward an earlier target-aligned structure
without returning to the same state.
Related work attributes nonmonotonic test error during training to modes learned at different
rates~\cite{advani2020high}. We instead track the learned function directly.

We study these dynamics by training a Transformer on cumulative-parity datasets
over binary strings of finite length $D$.
Because there are only a finite number of such strings, we evaluate each learned
Boolean output on all of them exactly. The sign table $h(x)=\pm 1$ defines the normalized function-state
\begin{equation}
  |\psi_h\rangle=2^{-D/2}\sum_x h(x)|x\rangle,
  \label{eq:function-state}
\end{equation}
where $x$ is the input binary string and $h(x)$ is the learned Boolean output.
Across a fixed bipartition $x=(a,b)$, where $a$ contains $d_L$ bits, fixing $a$
defines a conditional function $h_a(b)=h(a,b)$. We call $a$ and $b$ the left and right substrings, 
and write $A$ and $B$ for the corresponding complementary subsystems. Tracing out $b$
gives the reduced density matrix
$\rho_A=\Tr_B(|\psi_h\rangle\langle\psi_h|)$, which is the normalized Gram
matrix of pairwise overlaps among these conditional functions. Its spectral
decomposition is
\begin{equation}
  \rho_A=\sum_\alpha \lambda_\alpha
  |u_\alpha\rangle\langle u_\alpha|,
  \label{eq:rhoA-spectrum}
\end{equation}
where the orthonormal vectors $|u_\alpha\rangle$ are the left Schmidt modes of the learned function-state and $\lambda_\alpha$ are their weights. We
summarize this spectrum by the effective Schmidt rank
\begin{equation}
  \Kf=\frac{1}{\Tr(\rho_A^2)}
  =\frac{1}{\sum_\alpha\lambda_\alpha^2}.
  \label{eq:kf}
\end{equation}
Because $|\psi_h\rangle$ is pure, $\log\Kf$ is its
R\'enyi-2 entanglement across the input bipartition
\cite{laflorencie2016quantum}. In the
function-state representation, an increase in $\Kf$ means greater bipartite
R\'enyi-2 entanglement: $\rho_A$ becomes more mixed and its weight spreads
over a larger effective number of left Schmidt modes. In the underlying
classical sign table, the same change means that more independent
combinations of the conditional functions carry appreciable weight. We call this \emph{differentiation}. Conversely, a decrease in $\Kf$ means
reduced R\'enyi-2 entanglement and reconcentration onto fewer such modes. We call this \emph{recoherence}.

For any normalized $|u\rangle$ in $A$, we define
the mode population
\begin{equation}
  w(u)=\langle\psi_h|\left(|u\rangle\langle u|\otimes I_B\right)|\psi_h\rangle
  =\langle u|\rho_A|u\rangle
  \label{eq:mode-population}
\end{equation}
to be the weight the learned function-state places along $|u\rangle$ on the left
substrings. The exact parity target factorizes across the cut as
$h_\star(a,b)=s_a^\star t_b^\star$, and its normalized left target mode
\begin{equation}
  |u_\star\rangle=2^{-d_L/2}\sum_a s_a^\star|a\rangle
  \label{eq:target-mode}
\end{equation}
encodes the prescribed parity sign pattern over the left input substrings.
The target-mode population is $\wstar=w(u_\star)$. For the exact parity function,
$\rho_A^\star=|u_\star\rangle\langle u_\star|$ is a rank-one projector, so $\Kf^\star=1$ and $\wstar=1$.
Recoherence is \emph{target-selective} when
$\Kf$ decreases and $\wstar$ increases.

In contrast to the evolving learned function-state, the same dataset carries a constraint 
structure of its own. We construct one such graphical representation for each component 
of the vector output and describe one of them here. The rank-one
parity target assigns each observed pair $(a,b)$ a signed relation
\begin{equation}
  s_a^\star t_b^\star=y_{ab},
  \label{eq:parity-constraint}
\end{equation}
with $y_{ab}=\pm1$ the observed
label. The observed pairs therefore define a
signed bipartite graph whose vertices are the left and right input substrings
, and each edge between a left and a right vertex is signed by its
label. Within each connected component, the
relative target orientations are fixed up to a component-wide $\mathbb Z_2$
flip. Graph connectivity thus determines which relative signs are
identifiable from the data. Repeated data points reproduce existing constraints,
whereas new data points can add relations, merge connected components, and expand
the identifiable support. This graph depends on the input pairs and their
labels, independently of the function-state realized by the trained model.

The learned function-state and the data-defined constraint graph thus provide
two complementary structural descriptions. To determine how they evolve together and where
they decouple, we carry out two studies. One focuses on function-states following 
training trajectories through fitting, generalization onset, retreat, and
recovery.
The other keeps track of two branches that start from the same trained model and
receive equal amounts of either distinct or duplicated data. Within each branching study,
architecture, nominal dataset size, optimizer reset, update budget, and
minibatch schedule are held fixed, while the number and connectivity of
independent constraints differ. These studies reveal whether the
learned state can continue to reorganize after the constraint graph stops
changing. Retreat and recovery test whether that reorganization can reverse on
the same graph.

In the following, we present in detail in Sec.~\ref{sec:causal} how small additions of unseen data increase $\Kf$, whereas larger
additions decrease $\Kf$ and increase $\wstar$, even as
graph connectivity grows monotonically, which is shown in Sec.~\ref{sec:constraint-mechanism}. 
On a fixed, complete identifiable support, recoherence precedes target
alignment and predictive improvement. This is established in Sec.~\ref{sec:fixed-support-timing}. 
Section~\ref{sec:retreat} further shows that
one constraint structure can support multiple function-state organizations.
Learning therefore requires not
only constraint satisfaction but also dynamical selection on the available support.

\section{Function-State Framework}
\label{sec:formalism}

Having framed the general problem, we now turn to the concrete parity system
studied here. Below we elaborate the function-state map, its observables, and the graph
quantities used in the analysis.

\subsection{Exact Boolean-State Map}
\label{sec:boolean-map}

Let $x=(x_1,\ldots,x_D)\in\{0,1\}^D$ be the binary string we consider. The target bit at output position $j$ is
the running parity
\begin{equation}
  p_j(x)=x_1\oplus\cdots\oplus x_j,
  \label{eq:target}
\end{equation}
where $\oplus$ denotes addition modulo two. At a saved checkpoint, 
the sign of the network's output logit at a fixed output position $j$ defines a Boolean function
$h:\{0,1\}^D\to\{-1,+1\}$. In Secs.~\ref{sec:boolean-map}--\ref{sec:recoherence}, we suppress the output-position label
$j$ on all quantities derived from $h$. It is restored in
Sec.~\ref{sec:design} and resolved explicitly in
Sec.~\ref{sec:locality-frontiers}. Evaluating all $2^D$
strings gives the function-state
$|\psi\rangle\equiv|\psi_h\rangle$ in
Eq.~\eqref{eq:function-state}, with density operator
$\rho_{AB}=|\psi\rangle\langle\psi|$.
The map $h\mapsto|\psi\rangle$ is defined only up to a global sign, and every
quantity below is invariant under it. The full function-state is a pure, lossless representation
of the classical network's sign table.

\subsection{Reduced State, Schmidt Spectrum, and Differentiation}
\label{sec:reduced-state}

Partition the string after $d_L$ bits as $x=(a,b)$, where
$a=(x_1,\ldots,x_{d_L})$ and $b=(x_{d_L+1},\ldots,x_D)$, so that the
basis of input strings factorizes as $|x\rangle=|a\rangle_A|b\rangle_B$. 
We write the state as
\begin{equation}
 |\psi\rangle=\sum_\alpha s_\alpha
 |u_\alpha\rangle_A|v_\alpha\rangle_B,
 \qquad \lambda_\alpha=s_\alpha^2,
  \label{eq:schmidt}
\end{equation}
where $|u_\alpha\rangle_A$ and $|v_\alpha\rangle_B$ are orthonormal and the weights
$\lambda_\alpha$ of Eq.~\eqref{eq:rhoA-spectrum} are the squared Schmidt
coefficients. We call $\{\lambda_\alpha\}$ the Schmidt spectrum. Writing $u_\alpha(a)={}_A\langle a|u_\alpha\rangle$ and
$v_\alpha(b)={}_B\langle b|v_\alpha\rangle$, the individual outputs read
\begin{equation}
  h(a,b)=2^{D/2}\,{}_A\langle a|{}_B\langle b|\psi\rangle
  =2^{D/2}\sum_\alpha s_\alpha\,u_\alpha(a)\,v_\alpha(b),
  \label{eq:response-schmidt}
\end{equation}
which is the singular-value decomposition of the sign table.
Writing the sign table as $M(a,b)=h(a,b)$, the raw conditional Gram matrix is
\begin{equation}
  C_h(a,a')=2^{-(D-d_L)}\sum_b
  M(a,b)M(a',b)=2^{d_L}\sum_\alpha\lambda_\alpha\,u_\alpha(a)\,u_\alpha(a').
  \label{eq:gram}
\end{equation}
Reading each conditional function as a vector on $B$,
$|\phi_a\rangle=2^{-(D-d_L)/2}\sum_bM(a,b)|b\rangle$, makes
Eq.~\eqref{eq:gram} the matrix of inner products
$C_h(a,a')=\langle\phi_a|\phi_{a'}\rangle$.
Equivalently, tracing out $B$ produces the reduced density matrix  
\begin{equation}
  \rho_A=2^{-d_L}C_h,
  \label{eq:RDM}
\end{equation}
which is generally mixed.

Following standard density-matrix terminology, we call the off-diagonal elements 
of $\rho_A$, equivalently of $C_h$, coherences~\cite{cohentannoudji1977quantum}. Each is the overlap between two conditional functions.
Because $M=\pm1$, each $|\phi_a\rangle$ is normalized and every
diagonal element of $C_h$ equals one, so all model
dependence lies off the diagonal: large coherences make the conditional functions nearly parallel and concentrate the spectrum, whereas vanishing
coherences leave them orthogonal and $\rho_A$ maximally mixed.

Accordingly, the effective Schmidt rank $\Kf$ of Eq.~\eqref{eq:kf} is at least one, and an
increase in $\Kf$, or differentiation, corresponds to a broadened spectrum and
lower purity.

The singular values of $M$ are invariant under independent row and column sign
flips. We fix this freedom with the crossing gauge
\begin{equation}
 M_\times(a,b)=M(a,b)M(a,0_B)M(0_A,b)M(0_A,0_B).
  \label{eq:crossing-gauge}
\end{equation}
To make the gauge transformation explicit, define
\begin{equation*}
 \begin{aligned}
 D_A&=\operatorname{diag}_a M(a,0_B),&
 D_B&=\operatorname{diag}_b M(0_A,b),\\
 m_{00}&=M(0_A,0_B).&&
 \end{aligned}
\end{equation*}
Equation~\eqref{eq:crossing-gauge} then reads
\begin{equation*}
 M_\times=m_{00}D_AMD_B.
\end{equation*}
Since $D_A$ and $D_B$ are orthogonal sign matrices, with
$D_BD_B^\mathsf{T}=I$ and $m_{00}^2=1$, one can write
$\rho_A=2^{-D}MM^\mathsf{T}$ and
$\rho_A^\times=D_A\rho_AD_A^\mathsf{T}$. The spectrum, purity, and $\Kf$ are
preserved under this transformation. This convention ensures
$M_\times(a,0_B)=M_\times(0_A,b)=1$. Appendix~\ref{app:crossing-gauge} shows
that these conditions fix the sign freedom and that $M_\times$ is
itself unchanged by the flips.
The resulting Gram kernel is
\begin{equation}
  C_h^\times(a,a')=2^{-(D-d_L)}\sum_b
  M_\times(a,b)M_\times(a',b).
  \label{eq:crossing-gram}
\end{equation}


\subsection{Recoherence and Target Alignment}
\label{sec:recoherence}

Recoherence is a process with decreasing $\Kf$, meaning that the spectrum reconcentrates and purity
rises. The concentration can be in the target mode or in a different left mode. With $P_\star=|u_\star\rangle\langle u_\star|$, the target-mode population of
Eqs.~\eqref{eq:mode-population} and~\eqref{eq:target-mode} reads
$\wstar=\Tr(\rho_AP_\star)
=\sum_{\alpha}\lambda_{\alpha}
|\langle u_{\alpha}|u_\star\rangle|^2$ in the Schmidt basis.
Thus $\Kf$ measures spectral
concentration without specifying its direction, whereas $\wstar$ measures
population specifically in the task-relevant mode.
We therefore use the operational classification
\begin{equation}
 \begin{array}{ll}
 \text{differentiation:} & \Kf\uparrow,\\[2pt]
 \text{target-selective recoherence:} & \Kf\downarrow
 \ \text{and}\ \wstar\uparrow,
 \end{array}
 \label{eq:operational-definitions}
\end{equation}
where $\uparrow$ and $\downarrow$ mean that the quantity increases or decreases
between the two states being compared.

Our analysis below also uses the uniform-mode coherence
\begin{equation}
  \Gammone=2^{-2d_L}\sum_{a,a'}C_h^\times(a,a').
  \label{eq:gamma1}
\end{equation}
Let $|u_{uni}\rangle=2^{-d_L/2}\sum_a|a\rangle$ be the uniform left mode and
$|\phi_a^\times\rangle=2^{-(D-d_L)/2}\sum_bM_\times(a,b)|b\rangle$ the
normalized conditional function at fixed $a$. Then
\begin{equation}
  \Gammone=\langle u_{uni}|\rho_A^\times|u_{uni}\rangle
  =\left\|2^{-d_L}\sum_a|\phi_a^\times\rangle\right\|^2.
  \label{eq:gamma1-meaning}
\end{equation}
Thus $\Gammone$ measures how strongly the crossing-gauged conditional
functions share a common direction, the subscript recording that a single
direction is used. The gauge is set by the learned function, so this common direction is
itself learned rather than fixed by the parity target. We therefore use $\wstar$ for claims about target alignment. Appendix~\ref{app:crossing-gauge} explains why we fix the gauge this way, works out an example, and generalizes $\Gammone$ to $n$ directions.

\begin{figure}[tbp]
  \centering
  \includegraphics[width=0.99\linewidth]{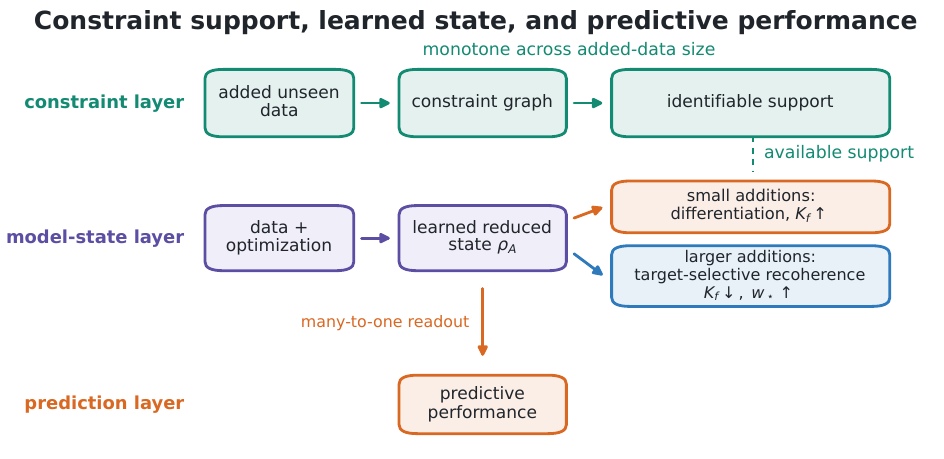}
  \caption{Identifiable support, model state, and predictive performance.
  Added unseen data monotonically expand the graph-defined identifiable support. Training on these data produces a
  reduced state within the available support. Across added-data size, small
  additions differentiate that state, whereas larger additions produce
  target-selective recoherence. Many different learned function-states yield the same predictive performance.}
  \label{fig:concept}
\end{figure}

\FloatBarrier

\section{Numerical Setup}
\label{sec:design}

With the observables of Sec.~\ref{sec:formalism} in mind, this section
specifies what we compute them on: which system we study, how position-resolved
quantities are aggregated into the reported ones, how predictive performance is
scored, and which training runs and contrasts each later section uses.

We study the case $D=11$, so the full binary space holds $2^{11}=2{,}048$
strings. All function-state observables are first computed separately for each output
position, with the fixed cut $d_L=5$.
Motivated by the output-position analysis in Sec.~\ref{sec:locality-frontiers}, the
primary analyses are position-averaged over the long-range outputs. The unindexed symbol denotes
\begin{equation*}
 O\equiv\frac{1}{4}\sum_{j=8}^{11}O^{(j)},
\end{equation*}
for a position-resolved observable $O^{(j)}$.
For example, $\Kf$ is the mean of the four
$\Kf^{(j)}$ values.

With $h^{(j)}$ and $h_\star^{(j)}$ being the learned and target 
sign functions at output $j$, both $\pm1$ valued, the block accuracy on a
held-out set $S$ of inputs is
\begin{equation}
  \Ablk(S)=\frac{1}{|S|}\sum_{x\in S}\prod_{j=1}^{D}
  \mathbf 1\bigl[h^{(j)}(x)=h_\star^{(j)}(x)\bigr],
  \label{eq:ablk}
\end{equation}
the fraction of inputs in $S$ whose $D$ output bits are simultaneously
correct. Held-out accuracies take $S$ to be the set of inputs never used for
training, that is, the complement of the training set within the 2,048-string
space. Training data are drawn with replacement. The $N=1280$ dataset
therefore covers 963 distinct strings and leaves $|S|=1{,}085$. The $N=64$
dataset leaves $|S|=1{,}984$. Full-space block accuracies instead evaluate Eq.~\eqref{eq:ablk} on all 2,048 strings. Given a threshold $\eta$, the generalization
onset of a trajectory is its first saved checkpoint with $\Ablk>\eta$. 
Throughout this work, we use $\eta=0.8$.  

The training runs and contrasts used below are summarized in Table~\ref{tab:designs}.
Of 30 trained $N=1280$ runs, we use the 24 that retreat by epoch 220.
Checkpoint grids,
optimizer resets, uncertainty calculations, and saved-checkpoint coverage are
given in Appendix~\ref{app:methods}.

\begin{table}[H]
\centering
\caption{Datasets, training spans, and corresponding analyses.}
\label{tab:designs}
\small
\setlength{\tabcolsep}{5pt}
\begin{tabular}{>{\raggedright\arraybackslash}p{0.19\linewidth}>{\raggedright\arraybackslash}p{0.38\linewidth}>{\raggedright\arraybackslash}p{0.35\linewidth}}
\toprule
analysis & data or model set and training span & checkpoints analyzed\newline and reported quantities \\
\midrule
small-data trajectories & $N=64$; 24 seeds; 4,000 epochs (= updates) & 52 checkpoints; fitting $\to$ interpolation \\
\midrule
large-data trajectories & $N=1280$; 24 seeds; 220 epochs & exact states from epoch 20 to generalization onset ($\Ablk>\eta$; epochs 51--168) \\
\midrule
state versus accuracy & $N=1280$; 24 seeds; 220 epochs & every checkpoint from epoch 20 to 220; $\Kf$, $\Gammone$ and $\wstar$ against full-space block accuracy \\
\midrule
branching study & 25 selected $N=64$ models ($\Kf\geq8.75$); additions $+16,+32,+64,+192,+576,+1{,}216$ per distinct branch; 800 updates each & distinct versus duplicate branches at 11 saved checkpoints for the $+1{,}216$ addition; endpoint distinct-minus-duplicate contrasts in $\Kf$ and $\wstar$ across all additions \\
\midrule
constraint graph & fixed $N=64$ base dataset; 10,000 nested addition sequences, each drawing 1,216 of the 1,984 unseen strings & connectivity probability and the identifiable fraction of left-substring pairs versus number of added data points \\
\midrule
retreat/recovery & first retreat in each of 24 seeds within the 220-epoch trajectories; 20 have a saved recovery & the last-high and retreat checkpoints, then the first recovery checkpoint where available \\
\bottomrule
\end{tabular}
\end{table}

Much of the analysis below, in
Secs.~\ref{sec:causal}--\ref{sec:fixed-support-timing}, rests on the branching
study, so we give its protocol in full here. Two branches are continued from one
selected starting model: its trained
weights are copied into both at a common branch point, and they differ only in
the data added. The distinct branch adds previously
unseen data points from the complement of the $N=64$ dataset, whereas the
duplicate branch enlarges that dataset by resampling data points already in it.
A third, control branch continues on the original dataset without any
addition. At the largest addition,
each augmented dataset contains 1,280 data points, the original 64 plus 1,216
additions. Every branch is initialized with AdamW's first- and
second-moment buffers set to zero, and uses learning rate $2\times10^{-3}$,
weight decay $10^{-4}$, batch size 128, 800 updates, and the same sequence of
random numbers for drawing minibatch indices.

In addition, the graph-only analysis keeps the original $N=64$ training set fixed and estimates
how connectivity changes as increasing numbers of randomly sampled,
previously unseen data points are added.

\section{Function-State Phenomenology}
\label{sec:observational}

Accordingly, our numerics have found the following phenomena, involving
timing, data dependence, and performance. Differentiation runs ahead of
prediction, reaching the longest-range outputs while those outputs are still
poorly predicted. Whether it then persists or gives way to recoherence depends
on how much data the model is given. And accuracy alone does not fix the state.
That is to say, function-states that predict equally well can be organized very
differently. Below we take these in turn.

\subsection{Output-Position Dependence}
\label{sec:locality-frontiers}

We take the position-resolved analysis first, since it justifies the
averaging over long-range outputs specified in Sec.~\ref{sec:design}. As output $j$ requires the parity
of the first $j$ input bits $x_1,\ldots,x_j$, increasing $j$
probes progressively longer-range dependence. We use this output axis to ask
whether function-state differentiation, target alignment, and
predictive performance reach the longer-range outputs together. 
Below, we consider 24 $N=1280$ trajectories of learning the $D=11$ system.
We measure training progress by the normalized time
$\tau=(e-20)/(e_{\rm gen}-20)$, where $e$ is the epoch and $e_{\rm gen}$ the
generalization-onset epoch of that seed, so that $\tau=0$ is epoch 20 and
$\tau=1$ is generalization onset. At 11 equally spaced values of $\tau$, we
then evaluate the model on the full binary space of 2,048 strings to obtain
the function-state observables and full-space accuracy.

\begin{figure}[tbp]
  \centering
  \includegraphics[width=0.99\linewidth]{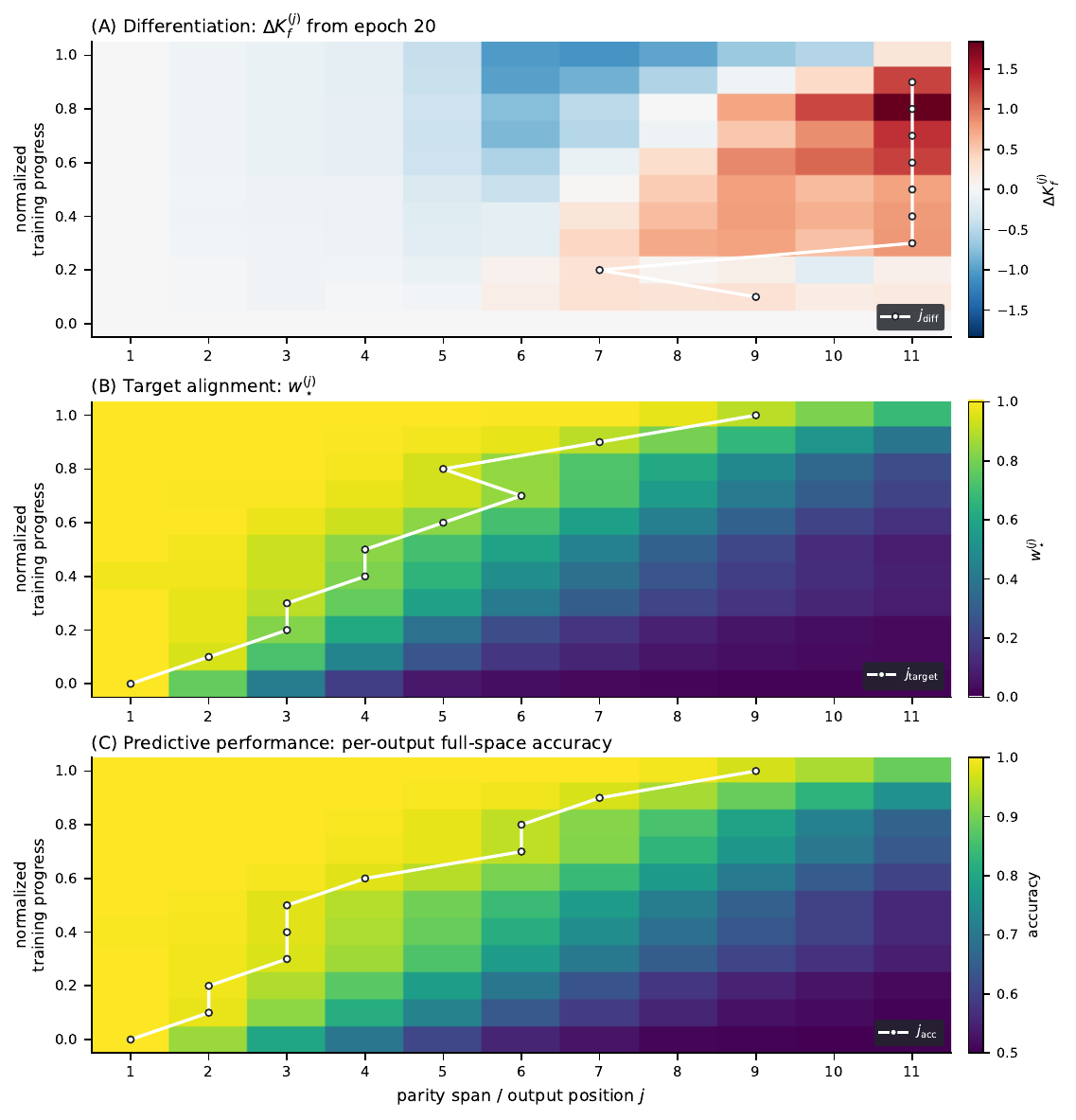}
  \caption{Output-position dependence from epoch 20 to generalization onset, averaged
  over 24 $N=1280$ runs. Rows are the 11 sampled checkpoints, and columns are
  outputs $j=1$--$11$. (A) Change in
  effective Schmidt rank relative to epoch
  20. (B) Target-mode population. (C) Accuracy over the full binary
  space. The overlaid lines mark the furthest outputs satisfying the
  differentiation, joint low-Schmidt-rank target-alignment, and accuracy criteria
  defined in the text. The lines guide the eye. The heatmap values are the
  primary quantities.}
  \label{fig:locality-frontiers}
\end{figure}

Fig.~\ref{fig:locality-frontiers} shows the averages across trajectories as
a function of training progress (rows) and output position $j$ (columns).
Panel A shows $\Delta\Kf^{(j)}=\Kf^{(j)}-\Kf^{(j)}\big|_{\text{epoch }20}$, the change in
effective Schmidt rank since epoch 20, and panels B and C show
$\wstar^{(j)}$ and the full-space accuracy. The line in each panel
marks the furthest output position meeting a fixed criterion at that
checkpoint. Here $j_{\rm diff}$ is the largest $j$ with
$\Delta\Kf^{(j)}\geq0.25$. Similarly, $j_{\rm target}$ is the largest $j$ whose state has
both low effective Schmidt rank and strong target alignment, as defined by
$\Kf^{(j)}\leq1.25$ and $\wstar^{(j)}\geq0.80$. Finally, $j_{\rm acc}$ is the
largest $j$ with accuracy at least $0.95$. 

Panel A shows almost no differentiation for $j\le d_L=5$ and clear structure from
$j=6$. Since the parity target factorizes as
$h_\star^{(j)}(a,b)=s_a^{\star,(j)}t_b^{\star,(j)}$ with
$t_b^{\star,(j)}\equiv1$ whenever $j\le d_L$, these outputs impose no
constraint across the cut. Their target state already contains only the single-mode, 
leaving differentiation no spectrum to broaden.

At normalized progress $0.3$, the differentiation criterion is satisfied
through $j=11$, although the furthest output position satisfying it changes
nonmonotonically over training. At generalization onset (normalized progress 1),
$j_{\rm target}=j_{\rm acc}=9$, so target alignment and high
accuracy have not yet reached outputs $j=10$ and 11. The function-state
can therefore continue to reorganize at long range before aligning
with the target and becoming accurate. We therefore average over $j=8$--$11$, hereafter the long-range
outputs, in the primary analyses. There the temporal separation between
differentiation and target alignment is clearest, allowing the two
processes to be resolved.

\FloatBarrier

\subsection{Small-Data Differentiation during Fitting}

\begin{figure}[tbp]
  \centering
  \includegraphics[width=0.99\linewidth]{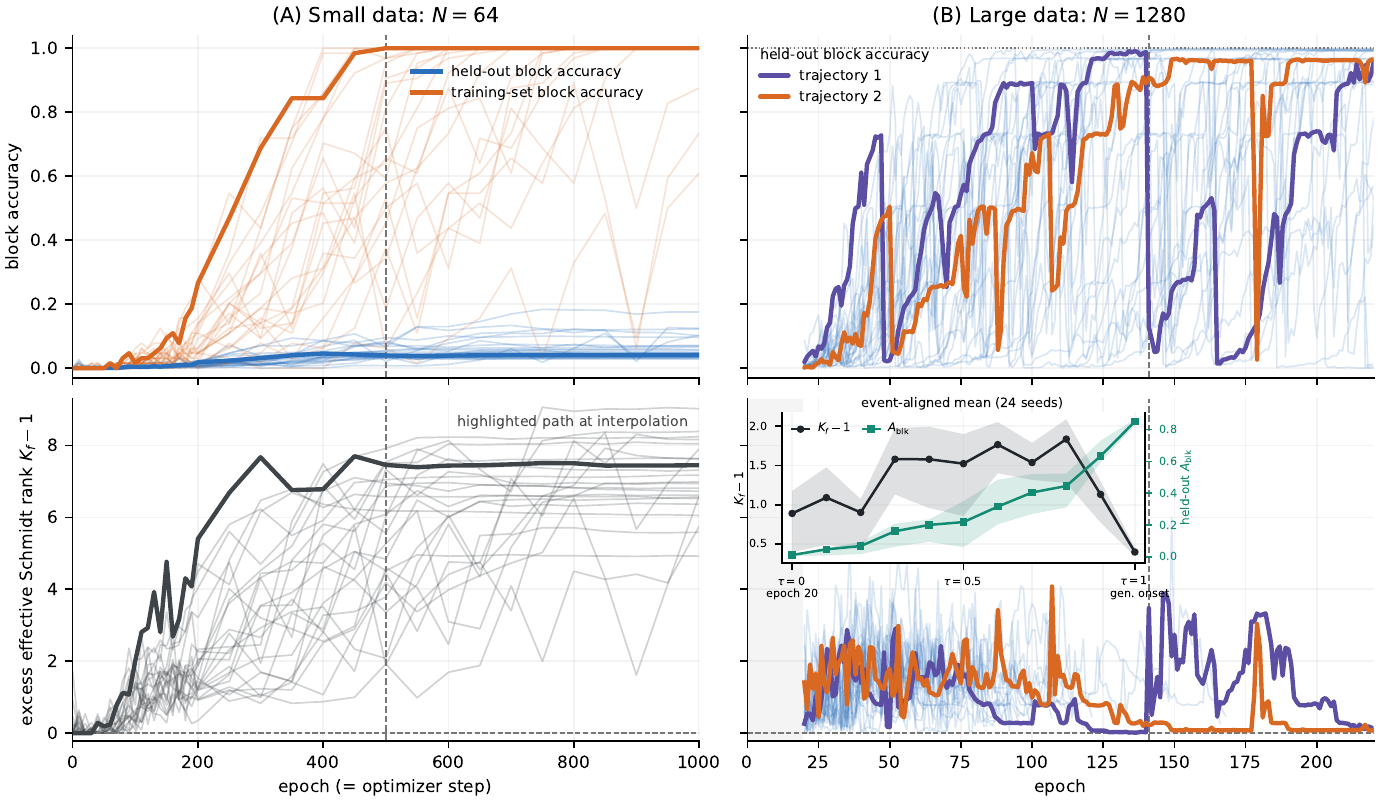}
  \caption{Training trajectories at small and large dataset sizes.
  Top: training-set and held-out block accuracy for $N=64$, and
  held-out block accuracy for $N=1280$. Bottom: excess effective Schmidt rank
  $\Kf-1$, whose single-mode value is zero. For $N=64$, interpolation leaves a
  persistent differentiated plateau. For $N=1280$, generalization onset is associated
  with low-$\Kf$ states. The highlighted retreat corresponds to a drop in accuracy
  and a rise in $\Kf-1$. For $N=1280$ both the accuracy and the saved exact
  function states begin at epoch 20, the first archived checkpoint. 
  Bold lines highlight one individual trajectory for
  $N=64$ and two for $N=1280$. Faint lines show the remaining seeds, and colors
  track the same $N=1280$ trajectories between the top and bottom panels. Dashed lines
  mark interpolation in the left panels and retreat in the right panels.
  The inset aligns epoch 20 to $\tau=0$ and each seed's generalization onset
  (epochs 51--168 across these trajectories) to $\tau=1$. Its two lines are
  the means of $\Kf-1$ and of held-out $\Ablk$ over the 24 seeds, each with
  shading spanning the interquartile range across seeds.}
  \label{fig:observational}
\end{figure}

We next ask how the function-state reorganizes as the model 
fits the training data, before substantial held-out generalization emerges. 
Fig.~\ref{fig:observational} follows the trajectories on their
training-epoch axes. We plot $\Kf-1$, which vanishes for a single Schmidt mode.
For $N=64$, fitting the training
dataset broadens the Schmidt spectrum. In each of the 24 seeds, interpolation
occurs at the first saved checkpoint at which
Eq.~\eqref{eq:ablk} equals one on the training set. 
At that checkpoint the medians across seeds are
$\Kf=8.00$, $\Gammone=0.097$, and held-out block accuracy
$\Ablk=0.050$. Interpolation is therefore reached with a
broad spectrum rather than a single-mode structure. The excess $\Kf-1$
rises toward 7 and remains high after interpolation, while held-out accuracy
remains poor.

\subsection{Large-Data Differentiation and Recoherence}

For $N=1280$, we test both stages in all 24 training trajectories, whose
generalization onset spans $e_{\rm gen}=51$--$168$ with median 94.
We therefore align them on the normalized time $\tau$, from epoch 20
at $\tau=0$ to generalization onset at $\tau=1$, and average at fixed
$\tau$ after linear interpolation in $\tau$. The mean $\Kf$ is
$1.89$ at $\tau=0$, reaches its largest value
$\max_\tau(\langle\Kf\rangle_{\rm seed})=2.97$ at $\tau=0.62$,
and falls to $1.40$ at $\tau=1$. 
In all 24 seeds, $\Kf$ exceeds its epoch-20 value at some later
checkpoint and then falls again by $\tau=1$. At these three times, mean
held-out block accuracy rises from $0.012$ to $0.339$ to $0.850$, with
both increases present in all seeds. The large-data trajectories therefore
pass through a higher-Schmidt-rank intermediate regime before recohering at
generalization onset.

A seedwise comparison further quantifies the recoherence
stage. Individual $\Kf$ peaks are spread over $\tau=0.01$--$0.88$ with median
$0.55$. Their mean $\langle\max_\tau\Kf\rangle_{\rm seed}=4.72$ exceeds
$\max_\tau(\langle\Kf\rangle_{\rm seed})$ because the peaks are not aligned. 
From those peaks to generalization onset, mean $\Kf$ falls to $1.40$,
while crossing-gauge $\Gammone$ rises from $0.186$ to $0.775$. Both changes have the stated direction in all
seeds. These changes establish spectral recoherence in the learned direction that
$\Gammone$ measures. Target-selective
recoherence is tested using $\wstar$ in Sec.~\ref{sec:causal}.

\subsection[State Multiplicity at Equal Accuracy]{State Multiplicity at Equal Accuracy}

\begin{figure}[tbp]
  \centering
  \includegraphics[width=0.99\linewidth]{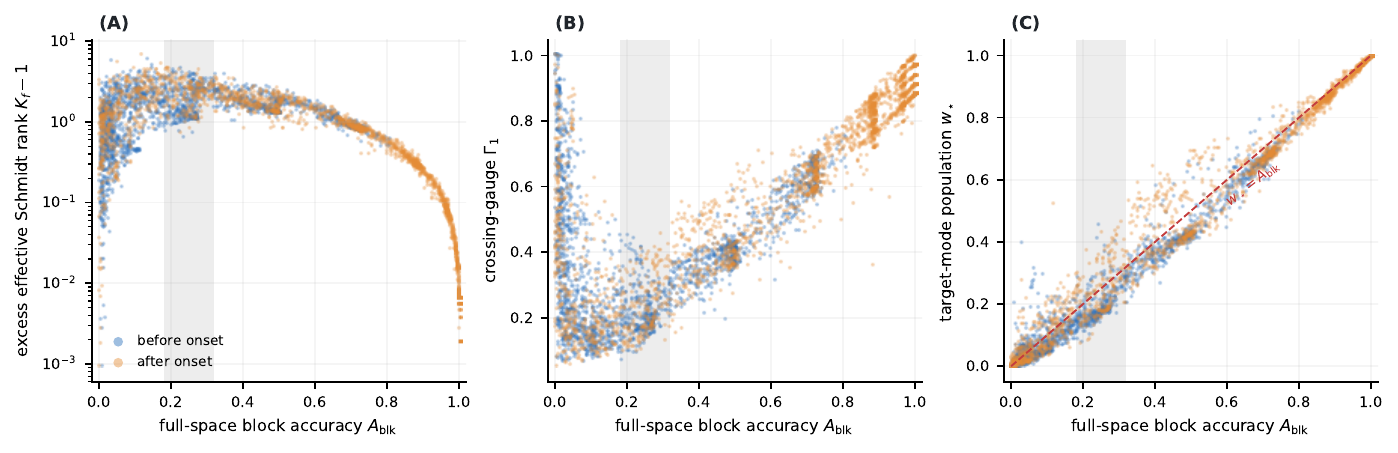}
  \caption{Function-state observables against predictive accuracy along the
  $N=1280$ trajectories. Each point is one checkpoint of the 24 runs from
  epoch 20 to epoch 220, so the range extends well past generalization onset.
  Blue marks checkpoints before onset and orange marks those at or after it.
  Gray shading marks the accuracy band
  $0.18\leq\Ablk\leq0.32$. (A) Excess effective Schmidt rank on a
  logarithmic scale. (B) Crossing-gauge $\Gammone$. (C) Target-mode population
  $\wstar$, with the dashed line marking $\wstar=\Ablk$. Accuracy here is
  evaluated on the full binary space.}
  \label{fig:state-vs-accuracy}
\end{figure}

To test whether accuracy determines the function-state, we compare
function-states that reach equal full-space accuracy in Fig.~\ref{fig:state-vs-accuracy}.
All three observables are multivalued at fixed accuracy. Among the
checkpoints with $0.18\leq\Ablk<0.22$, the excess effective Schmidt rank runs
from $0.75$ to $4.68$, crossing-gauge $\Gammone$ from $0.071$ to $0.454$ and
the target-mode population from $0.076$ to $0.373$. Each range is far wider
than the accuracy window that selects it. The ranges narrow as accuracy rises,
and by $0.88\leq\Ablk<0.92$ they have shrunk to $0.15$--$0.39$ for $\Kf-1$,
$0.716$--$0.925$ for $\Gammone$ and $0.829$--$0.925$ for $\wstar$.

The contrast survives in the limit of near-perfect prediction. Over the
checkpoints with $\Ablk>0.99$ the excess effective Schmidt rank is
confined below $0.024$ and the target-mode population to
$\wstar=0.988$--$0.999$, whereas $\Gammone$ still covers $0.847$--$0.999$.
That spread separates the runs rather than the checkpoints within a run.
Within one seed the standard deviation of $\Gammone$ over these checkpoints is
$0.0045$. Between the seed averages it is ten times larger at $0.043$. Each run
settles at its own value of $\Gammone$ and stays there. Accuracy therefore
does not uniquely specify the learned function-state. Models that are almost
indistinguishable in performance remain distinguishable as function-states.

\FloatBarrier
\section[Distinct-versus-Duplicate Data Branching Study]{Distinct-versus-Duplicate Data Branching Study}
\label{sec:causal}

The trajectories so far are observational. The amount of data and the
learned state vary together, so the reorganization can be correlated with the
added data but not attributed to it. We therefore intervene on the data rather
than observe it. One trained model is copied into branches that start
from identical weights and share the branch point, the reset optimizer
protocol, and the number of added data points. The branches differ only
in whether those points are previously unseen or resampled duplicates. A third
branch continues without any addition, since duplication still enlarges
the training set with data the model has already seen. Each branch is a
continuation of one and the same training run, so the differences that develop
between them separate the effect of new constraints carried by unseen data from 
that of duplication alone. Below we apply this branching
study (Table~\ref{tab:designs}) first at the largest addition and then across
added-data sizes.

\subsection[Function-State Recoherence at the Largest Addition]{Function-State Recoherence at the Largest Addition}

We first build the branches with the largest addition.
The branches continue a model already trained on the $N=64$ dataset. The
distinct branch adds 1,216 previously unseen input--output pairs and reaches
$N=1280$, whereas the duplicate branch adds the same number of resampled copies of data it has
already seen. Because the branches end up with
different training sets, every branch is evaluated on the same 2,048
inputs. The construction is repeated for 25 independently initialized runs
trained on the same fixed $N=64$ dataset and satisfying the
selection criterion $\Kf\geq8.75$. Across the 25 branch points the
mean values are $\Kf=9.398$ and $\Ablk=0.084$.

\begin{figure}[tbp]
  \centering
  \includegraphics[width=0.99\linewidth]{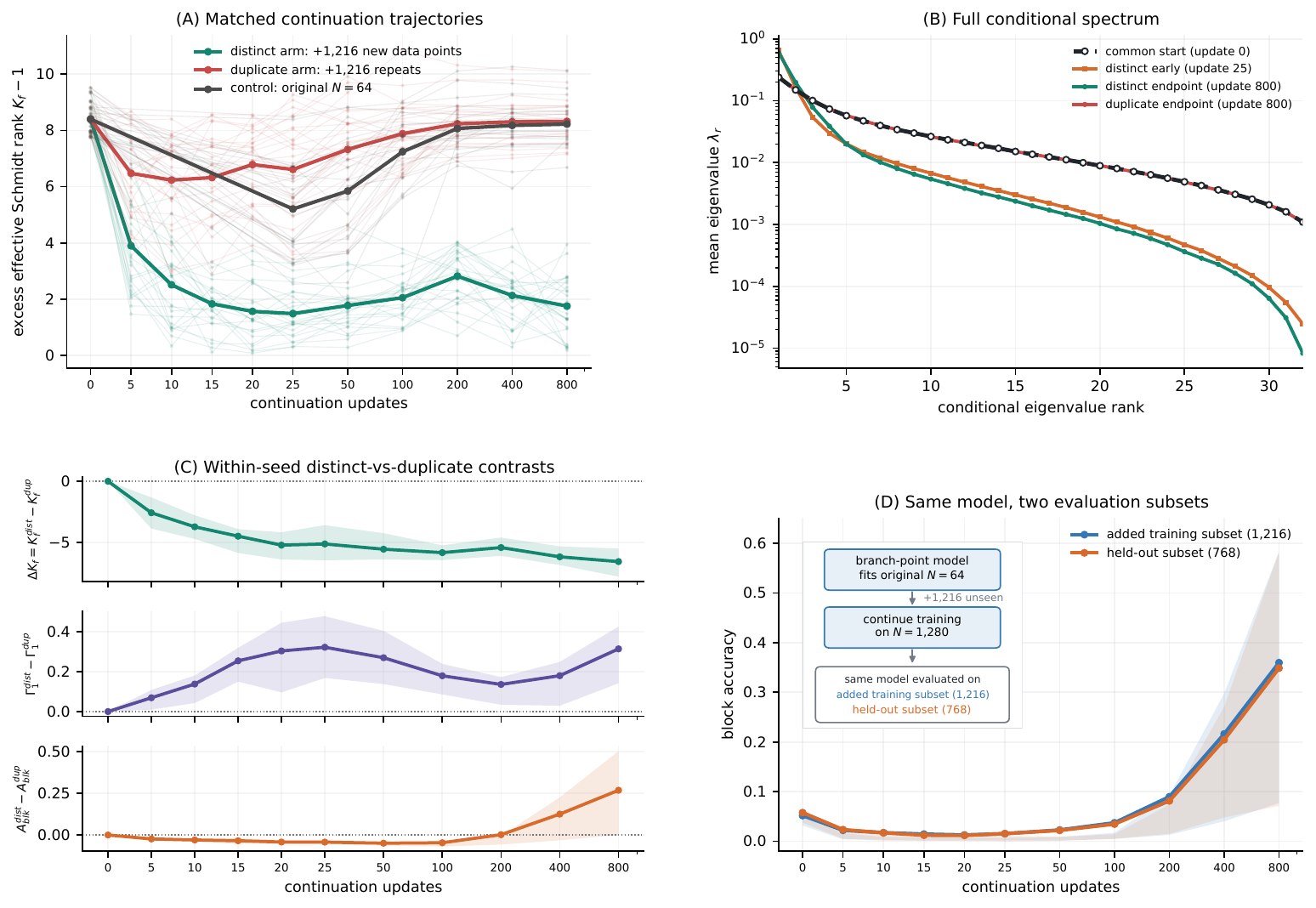}
  \caption{Branches from an identical starting model.
  (A) $\Kf-1$ along 800 continuation updates in the distinct, duplicate, and
  control branches, which respectively add 1,216 unseen input--output
  pairs, add the same number of resampled copies of the original data, and continue on the
  original $N=64$ training set. The distinct branch develops the lowest
  $\Kf-1$. (B) Mean Schmidt spectra at the common branch point, at the
  distinct-branch update-25 checkpoint, and at the two update-800
  endpoints. The common-start and duplicate-endpoint spectra nearly coincide,
  and the common start is drawn as a black dashed curve with open markers.
  By update 800, the distinct branch (green) has more weight in the leading
  Schmidt modes and less in the higher-rank tail than the common start and the
  duplicate endpoint (red).
  (C) Within-run distinct-minus-duplicate differences, averaged over 25 runs.
  The contrast $\Delta\Kf=\Kf^{\rm dist}-\Kf^{\rm dup}$ falls rapidly and
  stays negative. At update 25 the crossing-gauge contrast
  $\Gammone^{\rm dist}-\Gammone^{\rm dup}$ is positive while the block
  contrast is negative. (D) Block accuracy of the same
  distinct-branch model on two evaluation subsets, the 1,216 added data points
  in blue and the 768 never-trained inputs in orange. Both stay low at update 25
  and later improve together. The vertical inset summarizes the data addition
  and the two checkpoint evaluations. Lines and spectra show means across runs,
  with shaded interquartile ranges where shown.}
  \label{fig:causal}
\end{figure}

In Fig.~\ref{fig:causal}, the common branch point corresponds to update 0.
Panel A follows $\Kf-1$ in the three branches over the 800 continuation
updates. The distinct branch develops a lower $\Kf-1$ than both the
duplicate and the control branches.
At update 25, panel B shows that spectral weight in the distinct branch
has shifted from the higher-rank tail into the leading Schmidt modes relative to
the common branch point.
Panel C shows the within-run distinct-minus-duplicate branch differences along the continuation. After 25
updates, averaging these differences over the 25 runs gives
$\Delta\Kf=\Kf^{\rm dist}-\Kf^{\rm dup}=-5.119$ and
$\Gammone^{\rm dist}-\Gammone^{\rm dup}=0.322$, while the accuracy contrast
is $\Ablk^{\rm dist}-\Ablk^{\rm dup}=-0.043$. At the same checkpoint,
panel D evaluates the same distinct-branch model on two subsets and shows
that block accuracy remains low at $0.015$ on the 1,216 added data points and
$0.016$ on the 768 inputs the branch never trains on. The added points are
already in the training set, yet the branch predicts them no better than inputs
it has never seen. The reorganization in panel C therefore precedes any fit to
the new data rather than following from it.

Notably, $\Kf-1$ decreases immediately after the branch point, as shown in
panels A and C, whereas the ordinary $N=1280$
trajectories of Sec.~\ref{sec:observational} first differentiate and then
recohere. Their mean $\Kf$ in Fig.~\ref{fig:observational} rises
from $1.89$ at epoch 20 to $2.84$ at normalized progress $0.8$ and
then falls to $1.40$ at generalization onset.
The two phenomena are consistent because the branches start from the
highly differentiated state that training on the $N=64$ dataset leaves behind.
The subsequent decrease therefore captures
recoherence without a preceding post-branch differentiation stage.

\FloatBarrier
\subsection{Nonmonotonic Function-State Reorganization across Added-Data Size}
\label{sec:added-data-size}

At the same branch points, we now vary the size of the addition from $+16$ to
$+1{,}216$ previously unseen input--output pairs. Small additions differentiate
the state, whereas larger additions recohere it and increase the target-mode
population.
The distinct additions are nested: each larger addition contains all inputs 
used at smaller sizes, while every model is independently continued from the same 
$N=64$ branch point. We compare each distinct-data branch with the duplicate 
branch of equal size at update 800 and define both endpoint contrasts as distinct
minus duplicate,
\begin{equation}
 \Delta\Kf=\Kf^{\rm dist}-\Kf^{\rm dup},\qquad
 \Delta\wstar=\wstar^{\rm dist}-\wstar^{\rm dup}.
 \label{eq:added-data-contrasts}
\end{equation}
At additions $+16$, $+32$, and $+64$, the mean $\Delta\Kf$ values are positive and rise as
the added-data size increases, whereas at additions $+192$, $+576$, and $+1{,}216$,
they are negative with increasing magnitude. Small additions therefore broaden the
Schmidt spectrum, while larger additions reconcentrate it. The
corresponding mean $\Delta\wstar$ values are positive at all six additions and rise
by more than an order of magnitude from $+16$ to $+1{,}216$. The mean rises rapidly
up to $+192$ and more slowly at larger additions. This means that recoherence at the larger
additions is accompanied by increased target-mode population.

\begin{figure}[tbp]
  \centering
  \includegraphics[width=0.99\linewidth]{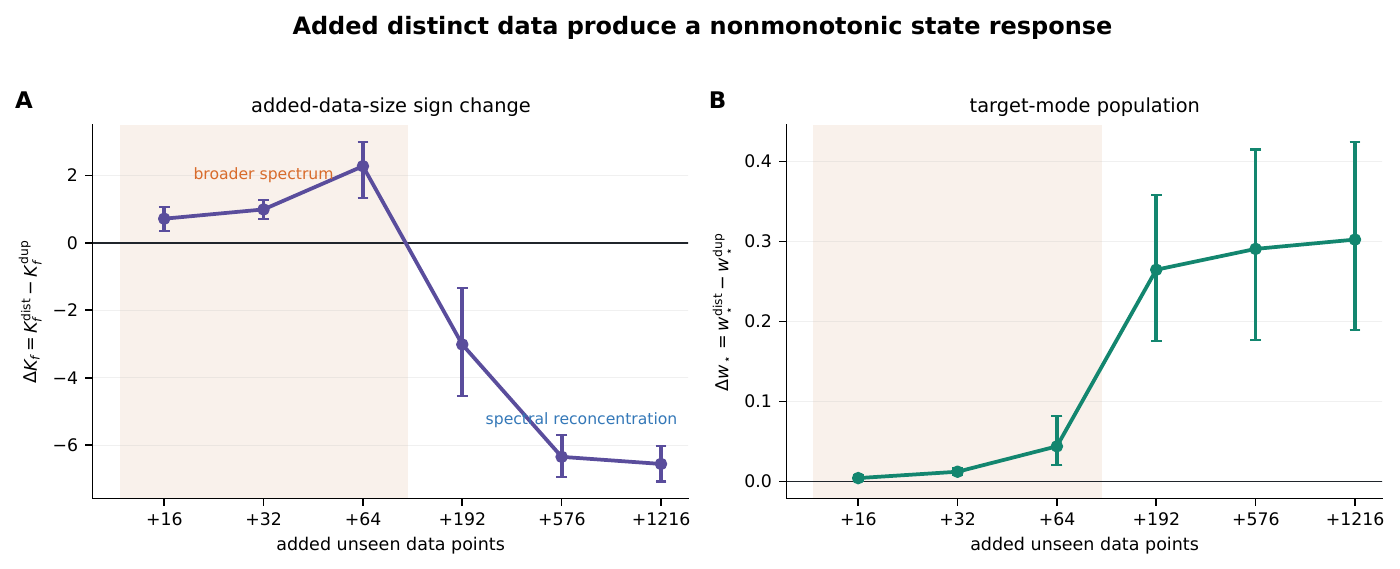}
  \caption{Nonmonotonic state reorganization across added-data size. Contrasts are
  distinct minus duplicate at update 800. Points are means across 25 
  starting models and bars are obtained in the way illustrated in Appendix~\ref{app:methods}. (A) Positive
  $\Delta\Kf$ at $+16,+32,+64$ marks differentiation. Negative $\Delta\Kf$ at
  $+192,+576,+1{,}216$ marks recoherence. (B) The mean of $\Delta\wstar$
  is always positive and grows across the series. Shading marks the three small data additions.}
  \label{fig:added-data-size}
\end{figure}

For the largest addition, the mean contrasts at update 800 are
$\Delta\Kf=-6.557$ and $\Delta\wstar=+0.302$. $\Delta\Kf$ is negative in
25 of the 25 starting models and $\Delta\wstar$ positive in 21 of the 25.
Their opposing signs 
show that the endpoint recoherence is target-selective. Later, in Sec.~\ref{sec:fixed-support-timing}, we will see
that the block 
accuracy advantage emerges only after
the state reorganization has persisted for hundreds of updates. An
independent-data-pool check is reported in Appendix~\ref{app:causal-support}.

\FloatBarrier
\section{Training-Data-Driven Bond Percolation on the Constraint Graph}
\label{sec:constraint-mechanism}

The interventions so far vary the number of added previously unseen data points,
but it is still unclear how those data constrain the target. Here we turn to the
structure of the training set itself, which fixes
identifiable relations of the target before any training. This structure is a property of
the data alone, whereas the Schmidt spectrum is a property of the learned
function. Following them together therefore separates the effect of the
data from the effect of the learning.

We represent the data as a graph. After splitting the input as 
$x=(a,b)$, we read the training
set as a bipartite constraint graph: the left vertices are the
substrings $a$, the right vertices are the substrings $b$, and an observed input
$x=(a,b)$ occupies the edge labeled by $e_{ab}\in\{-1,+1\}$ between them. 
As in a Tanner graph, edges
only exist between the two vertex sets.

The target has $D$ Boolean output positions, each of which corresponds to 
a signed graph $G_S^{(j)}$ with $j=1,\ldots,D$. The label
$y_{ab}^{(j)}\in\{-1,+1\}$ fixes $e_{ab}$ in $G_S^{(j)}$. These
graphs have the same connected components and differ only in their edge signs.
A duplicated data point repeats an occupied edge and leaves the graph
unchanged. A previously unseen input occupies a new edge and can merge two
connected components. Accumulating distinct data is therefore bond percolation on the
bipartite constraint graph.

The connectivity statistics below are sign-insensitive and therefore the same
for all $D$ outputs. They record which relative target signs the data make
identifiable.

\subsection{Graph Observables and Their Calculation}

We first ask how much of the target the data alone can fix.
For each parity target, the edge labels determine all relative target
signs between any two vertices within each connected component, 
up to one component-wide flip. Let
$c(G_S)$ denote the number of components, including isolated vertices. After
removing one global sign, the number of unresolved relative orientations is
\begin{equation}
  \dfree=c(G_S)-1.
  \label{eq:dfree}
\end{equation}
Full connectivity is therefore the exact criterion for 
sign identifiability across all substrings.

To quantify the identifiable support on the left side of the cut, define the
connectivity selector
\begin{equation}
 J_{aa'}=
 \begin{cases}
 1, & a\text{ and }a'\text{ lie in the same graph component},\\
 0, & \text{otherwise}.
 \end{cases}
 \label{eq:left-connectivity-selector}
\end{equation}
The right substrings remain part of the full bipartite graph and mediate
connectivity between left substrings.

Fixed by the data alone, $J$ also defines the unit-trace reference state
\begin{equation}
 \rho_{\rm perc}=\frac{J}{n_L},
 \label{eq:rhoperc-J}
\end{equation}
which we use below to compare graph connectivity with the learned reduced
state. Here we define $n_L=|V_L|=2^{d_L}$ as the dimension of the full left space, and $n_{L,c}=|c\cap V_L|$ as the number of left
vertices in component $c$. Let
$p_c=n_{L,c}/n_L$ and associate each component containing left substrings with
the normalized component-coherent mode
\begin{equation*}
 |u_c^G\rangle=\frac{1}{\sqrt{n_{L,c}}}
 \sum_{a\in c\cap V_L}|a\rangle.
\end{equation*}
We can write $\rho_{\rm perc}=\sum_{c:n_{L,c}>0}p_c|u_c^G\rangle\langle u_c^G|$.
This is Eq.~\eqref{eq:rhoperc-J} expressed using quantum-information notation.
This graph-only reference state $\rho_{\rm perc}$ is fully coherent within each
left-substring component and incoherent between different components. It occupies
every graph-identifiable left-substring relation with equal magnitude. Changing the
within-component sign convention only conjugates $\rho_{\rm perc}$ by a
diagonal unitary and leaves its spectrum unchanged. We define $K_{\rm perc}$
as the inverse purity of $\rho_{\rm perc}$, in parallel with $\Kf^{(j)}$ for $\rho_A$. 
Since the component modes are
orthonormal, direct computation gives
\begin{equation}
 \begin{aligned}
 K_{\rm perc}^{-1}
 &\equiv \Tr\rho_{\rm perc}^2
 =\sum_{c\in\mathcal C(G_S)}p_c^2
 =\frac{1}{n_L^2}\sum_{c\in\mathcal C(G_S)}n_{L,c}^2\\
 &=\frac1{n_L}+\frac{1}{n_L^2}
   \sum_{c\in\mathcal C(G_S)}n_{L,c}(n_{L,c}-1)\\
 &=\frac1{n_L}+\frac{n_L-1}{n_L}f_L
  =\frac1{32}+\frac{31}{32}f_L,
 \end{aligned}
 \label{eq:kperc-null}
\end{equation}
where the last equality uses $n_L=2^{d_L}=32$. $K_{\rm perc}$ is the effective 
number of left-substring components weighted by their sizes.
 Equation~\eqref{eq:kperc-null}
is analytic once the component sizes, or equivalently $f_L$, are specified.

The ratio $f_L$ appearing there is the identifiable fraction of ordered
off-diagonal left-substring pairs,
\begin{equation}
 f_L=\frac{1}{n_L(n_L-1)}\sum_{a\ne a'}J_{aa'}
 =\frac{\sum_{c\in\mathcal C(G_S)}n_{L,c}(n_{L,c}-1)}
 {n_L(n_L-1)}.
 \label{eq:fL}
\end{equation}
Accordingly, $f_L$ is the fraction of left-substring pairs in the
same component of the full bipartite graph. A path through the training edges then fixes their relative
target sign. No such relation is identifiable between different components.
Thus $c(G_S)$ and $\dfree$ describe connectivity across the full graph,
whereas $f_L$ measures the identifiable fraction of left-substring pairs.
Since $f_L$ cannot decrease when edges are added, $K_{\rm perc}$ decreases
monotonically toward one.

\begin{figure}[tbp]
  \centering
  \includegraphics[width=0.99\linewidth]{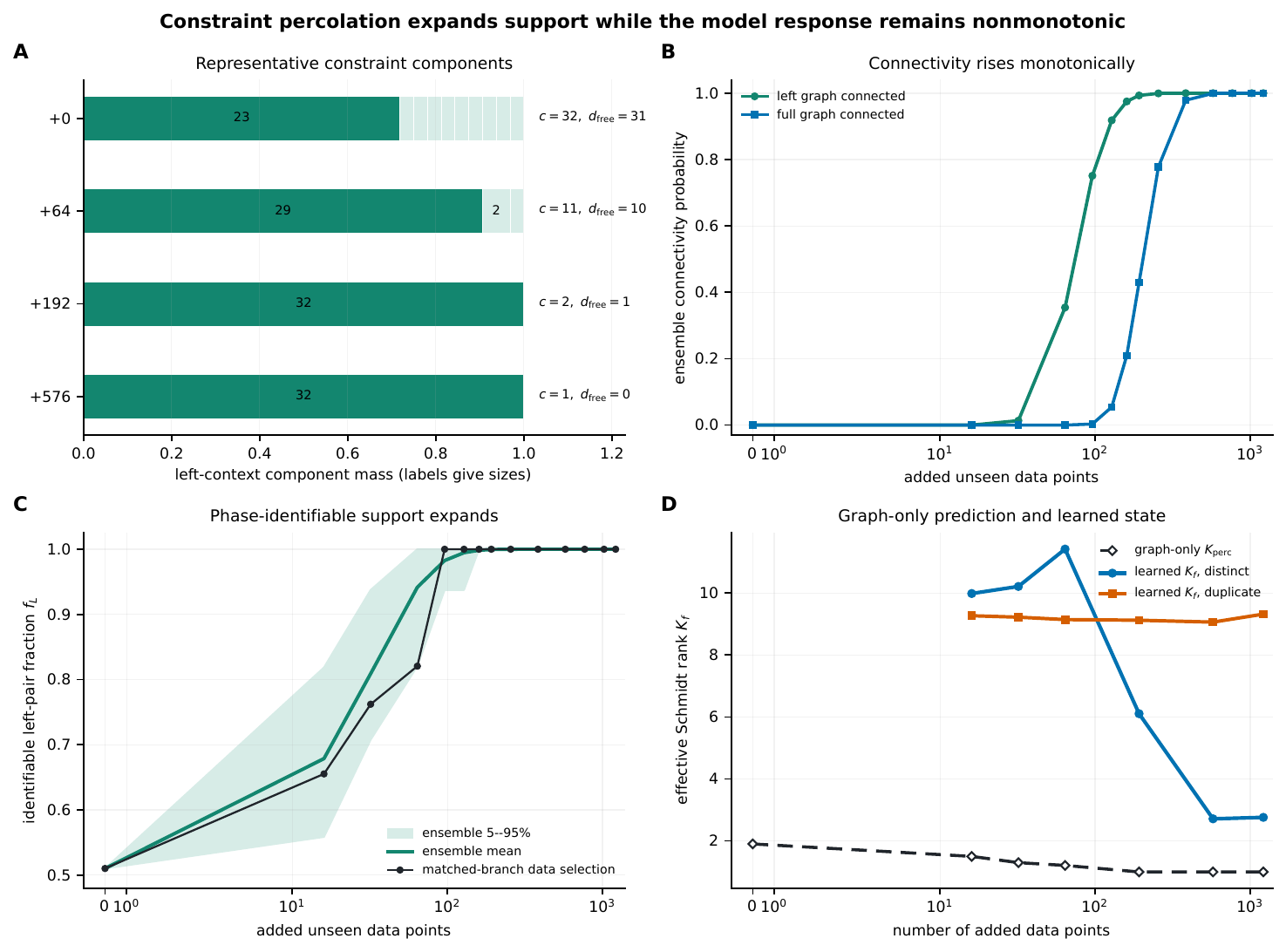}
  \caption{Bond percolation on the constraint graph and comparison with the
  learned function-state. (A) Exact component decompositions for the nested
  datasets $+0$, $+64$, $+192$, and $+576$ used in the branching study of Sec.~\ref{sec:added-data-size}. Bar segments give the fractions of the 32 left substrings in
  each full-graph component, and segment labels give their counts. Among all
  96 vertices, $c$ is the number of connected components and
  $\dfree=c-1$ counts the independent component signs that remain undetermined.
  (B) Connectivity probabilities over 10,000 random cumulative additions to
  the same $N=64$ graph. Each realization places the 1,984 previously unseen inputs in a
  random order and adds the first 1,216. The $+16$ dataset uses the first 16. The $+32$ dataset retains those
  inputs and adds the next 16. Larger datasets are constructed in the same way
  up to $+1{,}216$.
  (C) The exact $f_L$ values for the same branching-study data selection, together with the mean and 5--95\% range from the
  same realizations. (D) The analytic $K_{\rm perc}$ from
  Eq.~\eqref{eq:kperc-null} and the update-800 learned $\Kf$ for the
  distinct and duplicate branches. The learned curves are means over 25
  starting models.}
  \label{fig:constraint}
\end{figure}

\subsection{Monotone Graph Connectivity and Nonmonotonic State Reorganization}

Fig.~\ref{fig:constraint} evaluates these graph observables across
added-data size and compares them with the learned Schmidt rank.
As distinct data are added, constraint graph connectivity and identifiable
left-substring support increase monotonically, whereas the learned Schmidt
rank changes nonmonotonically.

For the branching-study data selection, Fig.~\ref{fig:constraint}A shows how
the graph components merge as distinct inputs are added. Each bar partitions
the 32 left substrings according to the full-graph component to which they
belong. Each segment represents one such component, with width proportional
to its number of left substrings. The displayed numbers give these counts,
whereas $c$ and $\dfree=c-1$ are computed from all 96 vertices. By $+576$, all left and right
substrings belong to a single component, so $\dfree=0$.

Fig.~\ref{fig:constraint}B asks how connectivity depends on the composition
of the added dataset. We first generate 10,000 random data-addition
realizations. For each realization, the 1,984 inputs absent from the original $N=64$
training set are placed in a random order, and the first 1,216 of them are
added in sequence. At $+16$, the first 16 inputs are added. At $+32$,
those 16 are retained and the next 16 are added. At $+64$, the 32 already
present are retained and the next 32 are added. The remaining sizes are constructed in the same cumulative
way up to $+1{,}216$. Thus the datasets are nested within each realization,
while the selected subset at any fixed size differs across realizations. At
each size, the left-connectivity curve is the fraction of realizations in
which all 32 left substrings lie in one component, and the full-connectivity
curve is the fraction in which all 96 vertices lie in one component. The
left-connectivity probability reaches
one half near $+76$ additions, whereas the full-connectivity probability does
so near $+205$.

The connectivity indicators in Fig.~\ref{fig:constraint}B are all-or-none.
Fig.~\ref{fig:constraint}C instead measures the fraction of
left-substring pairs that lie in the same connected component.
Equation~\eqref{eq:fL} is evaluated for each of the 10,000 realizations of
panel B. The black curve shows $f_L$ for the fixed branching-study
data selection. At each added-data size, the colored curve is the mean $f_L$
over the 10,000 random realizations, and the shaded band spans the 5th--95th
percentiles. For the fixed selection, $f_L$ reaches one
by $+96$ before the full graph is connected. Since $f_L$ tests only whether
left-substring pairs share a component, right substrings can remain disconnected
without lowering it.

Fig.~\ref{fig:constraint}D then asks whether this monotone growth of
identifiable support determines the learned Schmidt rank. The dashed
curve evaluates $K_{\rm perc}$ for the idealized graph state $\rho_{\rm perc}$
using Eq.~\eqref{eq:kperc-null}. It decreases monotonically to one as $f_L$
increases to one. The learned $\Kf$ in the distinct-data branch instead first
rises and then falls, while in the duplicate branch it remains near its starting value. The identifiable support therefore grows monotonically with the data, while the
learned Schmidt spectrum does not follow it. Section~\ref{sec:fixed-support-timing} sharpens this
separation by holding the graph fixed while the learned function continues to
reorganize.

\FloatBarrier
\section[Learned-State Reorganization on Fixed Identifiable Support]{Learned-State Reorganization on Fixed Identifiable Support}
\label{sec:fixed-support-timing}

We now hold the identifiable support fixed and ask when correlation magnitude, target
alignment, and predictive accuracy change as learning continues on a fixed
training set. The $+1{,}216$ distinct-data branch provides the limiting case:
$f_L=1$ at update 0 and remains one thereafter, as shown in
Fig.~\ref{fig:constraint}, so every relative target sign between left substrings
is identifiable from the first update onward. Any later change in the
correlations, target alignment, or accuracy of the learned function therefore
cannot be attributed to the constraint graph. We use
this distinct-data branch, compared with its duplicate branch, to place those changes in
time.

\subsection{Correlation Magnitude Precedes Target Alignment}
\label{sec:graph-dephasing}

As training proceeds, a pair of left substrings can change in two ways.
The learned function can tie the pair together, making $|C_h(a,a')|$ large. It can also align that tie with the target, so that the sign of $C_h(a,a')$
agrees with $s_a^\star s_{a'}^\star$. These are independent properties of the same
off-diagonal element, and a pair can become strongly correlated well before it
becomes correctly signed. Purity responds to the magnitude alone, whereas the
target-mode population rewards only the correctly signed part.

To order these changes in time, we track two averages over pairs of left substrings,
\begin{equation}
 \begin{aligned}
 Q_{\rm in/out}^{(j)}&=\mathbb E[(C_h^{(j)}(a,a'))^2\mid J_{aa'}=1/0],\\
 R_{\rm in/out}^{(j)}&=\mathbb E[s_a^{\star,(j)}s_{a'}^{\star,(j)}C_h^{(j)}(a,a')\mid
 J_{aa'}=1/0],
 \end{aligned}
 \label{eq:pair-realization}
\end{equation}
where the selector $J_{aa'}$ is defined in Eq.~\eqref{eq:left-connectivity-selector}, and the $1/0$ two cases are labeled ``in'' and ``out.''
The first, $Q$, averages the squared
correlation $C_h(a,a')^2$ and measures how strongly the learned function ties
two left substrings together, irrespective of sign. The second, $R$, averages the
target-signed correlation $s_a^\star s_{a'}^\star C_h(a,a')$. Positive values
indicate target-aligned correlations, whereas negative values indicate
target-opposed correlations. For both \(Q\) and \(R\), we compute one 
average over identifiable pairs (\(J_{aa'}=1\)) and another over 
unidentifiable pairs (\(J_{aa'}=0\)). Graph connectivity determines this 
partition. Purity and target-mode population then
decompose as
\begin{equation}
 [\Kf^{(j)}]^{-1}=\Tr[(\rho_A^{(j)})^2]
 =\frac{1}{n_L^2}\sum_{a,a'}[C_h^{(j)}(a,a')]^2
 =\frac1{n_L}+\frac{n_L-1}{n_L}
 [f_LQ_{\rm in}^{(j)}+(1-f_L)Q_{\rm out}^{(j)}],
 \label{eq:kf-factorized}
\end{equation}
\begin{equation}
 \wstar^{(j)}=\langle u_\star^{(j)}|\rho_A^{(j)}|u_\star^{(j)}\rangle
 =\frac{1}{n_L^2}\sum_{a,a'}s_a^{\star,(j)}s_{a'}^{\star,(j)}C_h^{(j)}(a,a')
 =\frac1{n_L}+\frac{n_L-1}{n_L}
 [f_LR_{\rm in}^{(j)}+(1-f_L)R_{\rm out}^{(j)}].
 \label{eq:w-factorized}
\end{equation}
For the fixed $d_L=5$ cut used in the primary analyses, $n_L=32$, so the
coefficients are $1/32$ and $31/32$. The graph fixes $f_L$, while
$Q_{\rm in/out}^{(j)}$ and $R_{\rm in/out}^{(j)}$ describe how the learned function uses
that available support. Equations~\eqref{eq:kf-factorized}
and~\eqref{eq:w-factorized} therefore separate constraint identifiability from
the state realized by training.

So far the in and out averages only sort the pairs into two groups. We now
realize the same split as a map acting on $\rho_A$ itself, which turns
$Q_{\rm in/out}$ and $R_{\rm in/out}$ into separate contributions to purity and
target-mode population.
For a fixed output position \(j\), we suppress the superscript \((j)\) on every quantity that depends on it. Let
$\Pi_c=\sum_{a\in c\cap V_L}|a\rangle\langle a|$ be the projector onto the left
substrings in component $c$ of the unsigned constraint graph, which is independent of the output position. We define the graph-component dephasing channel as
\begin{equation}
 \mathcal D_G(\rho_A)=\sum_c\Pi_c\rho_A\Pi_c.
 \label{eq:graph-dephasing}
\end{equation}
This channel retains \(\rho_A(a,a')\) when \(J_{aa'}=1\) and removes it when
\(J_{aa'}=0\).

Because the $\Pi_c$ are mutually orthogonal and sum to the identity,
Eq.~\eqref{eq:graph-dephasing} is the standard pinching channel, or
equivalently the non-selective update obtained by measuring the
component label and discarding the outcome
\cite{cohentannoudji1977quantum}. It is completely positive and
trace-preserving. Because the projectors can have rank
larger than one, this is a block dephasing: it erases coherence between
different components while preserving coherence within each component. It can
therefore be viewed as the complete-decoherence limit of an environment that
monitors only the component label \cite{zurek2003decoherence}.

The same projector-sum structure appears in stabilizer quantum error
correction~\cite{gottesman1997stabilizer}. Joint measurement of commuting parity checks partitions the
physical Hilbert space into syndrome sectors $\{P_s\}$. Discarding the
syndrome record gives the same channel form,
$\rho\mapsto\sum_sP_s\rho P_s$. As in the present setting, the syndrome outcomes do not fix the
logical state within a sector. Thus the component label
$c$ here plays a role formally analogous to a syndrome label $s$. This structural analogy allows us to use the connected components induced by the training-data graph to decompose learned-state coherence into within- and between-component contributions.

The channel gives a direct quantum interpretation of the pair averages in
Eq.~\eqref{eq:pair-realization}. Let
$w_{\star,G}=\langle u_\star|\mathcal D_G(\rho_A)|u_\star\rangle$. Because
$\mathcal D_G$ retains the within-component blocks and removes the
between-component blocks, Eq.~\eqref{eq:graph-dephasing} gives
\begin{equation}
 \begin{aligned}
 \Tr[\mathcal D_G(\rho_A)^2]
 &=\frac1{n_L}+\frac{n_L-1}{n_L}f_LQ_{\rm in},\\
 \Tr(\rho_A^2)-\Tr[\mathcal D_G(\rho_A)^2]
 &=\frac{n_L-1}{n_L}(1-f_L)Q_{\rm out},\\
 w_{\star,G}
 &=\frac1{n_L}+\frac{n_L-1}{n_L}f_LR_{\rm in},\\
 \wstar-w_{\star,G}
 &=\frac{n_L-1}{n_L}(1-f_L)R_{\rm out}.
 \end{aligned}
 \label{eq:channel-pair-decomposition}
\end{equation}
Thus $Q_{\rm in/out}$ determine the within- and between-component contributions
to purity, while $R_{\rm in/out}$ determine the corresponding contributions
to target-mode population. The $Q$ contributions are
nonnegative, whereas an $R$ contribution can be target-aligned or
target-opposed. If there are no within-component or no between-component pairs, the corresponding conditional average is undefined. Its contribution to Eq.~\eqref{eq:channel-pair-decomposition} is nevertheless zero because the corresponding pair fraction vanishes.

\begin{figure}[tbp]
  \centering
  \includegraphics[width=0.92\linewidth]{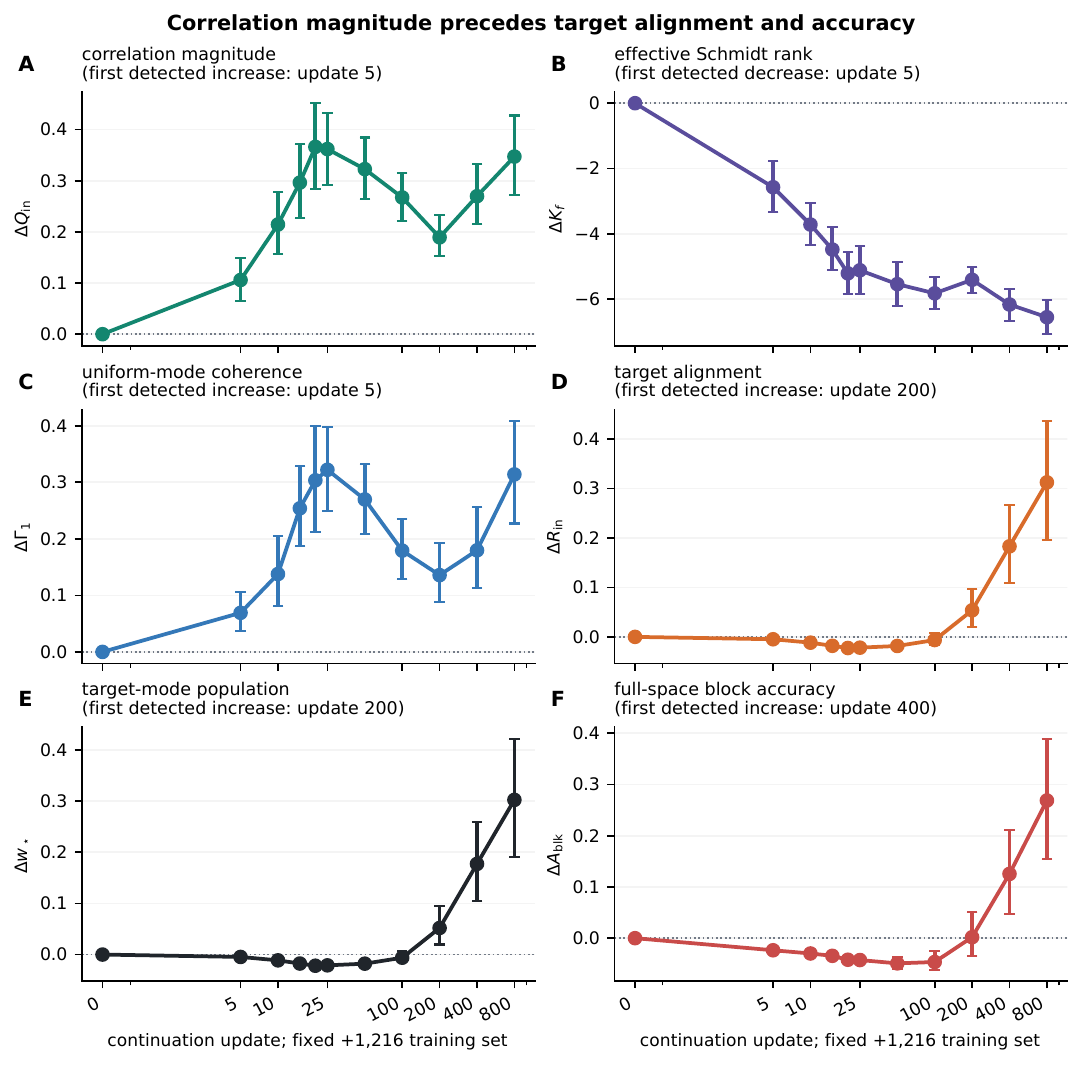}
  \caption{Learned-state and accuracy changes after the identifiable support is complete
  in the $+1{,}216$ distinct-data branch, where $f_L=1$ from update 0. Each point
  is the mean distinct-minus-duplicate difference across 25 starting
  models, and error bars are 95\% bootstrap intervals for that mean (Appendix~\ref{app:methods}). The
  state quantities are averaged over $j=8$--$11$. Here $\Delta\Ablk$ is the
  full-space block accuracy contrast. ``First detected increase'' or
  ``decrease'' denotes the first saved checkpoint at which the 95\% interval
  lies entirely above or below zero, respectively. This occurs for
  $\Delta Q_{\rm in}>0$, $\Delta\Kf<0$, and $\Delta\Gammone>0$ at update 5. It occurs for $\Delta R_{\rm in}>0$ and $\Delta\wstar>0$ at update 200 and for $\Delta\Ablk>0$ at update 400.}
  \label{fig:fixed-support-time}
\end{figure}

For the $+1{,}216$ distinct-data branch, $f_L=1$ from update 0, so all
left-substring pairs belong to the same component. The component projector on
$A$ is therefore the identity, and the channel acts trivially:
$\mathcal D_G(\rho_A)=\rho_A$. For each output position and branch,
$Q_{\rm in}=(n_L/\Kf-1)/(n_L-1)$ and
$R_{\rm in}=(n_L\wstar-1)/(n_L-1)$. These are linear functions of the purity $1/\Kf$ and of $\wstar$, respectively. After averaging over $j=8$--$11$ in
each branch, we define the paired contrasts at each saved update
as $\Delta Q_{\rm in}=Q_{\rm in}^{\rm dist}-Q_{\rm in}^{\rm dup}$ and
$\Delta R_{\rm in}=R_{\rm in}^{\rm dist}-R_{\rm in}^{\rm dup}$. We average
them over the 25 starting models. The results are in
Fig.~\ref{fig:fixed-support-time}A and D. They show that the error-bar regions
first become entirely positive at update 5 for $\Delta Q_{\rm in}$ and at update 200 for
$\Delta R_{\rm in}$. The error-bar calculation follows
Appendix~\ref{app:methods}. The purity increase is therefore
detected before the increase in target-mode population. The state observables introduced above show the same temporal separation: $\Delta\Kf<0$ and
$\Delta\Gammone>0$ are first detected at update 5, whereas
$\Delta\wstar>0$ is first detected at update 200
(Fig.~\ref{fig:fixed-support-time}B, C, and E). Thus $\Gammone$ follows the
early recoherence, while $\wstar$ follows the later alignment with the fixed
target mode.

\subsection{Predictive Improvement Occurs Later}

We also have the block contrast $\Delta\Ablk$ in
Fig.~\ref{fig:fixed-support-time}F. Its error-bar region first becomes
entirely positive at update 400. Complete identifiable support is therefore available
before the model realizes the corresponding correlations, and the resulting
state organization becomes detectable in predictive accuracy only later. On
the saved checkpoint grid, the observed ordering is
\begin{equation}
  \boxed{\begin{gathered}
  \text{state differentiated at }N=64,\ \text{support completed}\ \to\\
  \{\Delta Q_{\rm in}>0,\ \Delta\Kf<0,\ \Delta\Gammone>0\}_{t=5}
  \ \to\\
  \{\Delta R_{\rm in}>0,\ \Delta\wstar>0\}_{t=200}
  \ \to\ \{\Delta\Ablk>0\}_{t=400}.
  \end{gathered}}
  \label{eq:timing}
\end{equation}
The first line gives the initial conditions of the branch. The arrows mark the
first detected changes in the directions shown. The
quantities $Q_{\rm in}$, $\Kf$, and $\Gammone$ track the early recoherence,
whereas $R_{\rm in}$ and $\wstar$ track the later target alignment.

\FloatBarrier
\section{Retreat and Recovery}
\label{sec:retreat}

Holding the identifiable support fixed does not stop the learned state from
reorganizing. We now ask whether such
reorganization can also reverse during ordinary training.
Retreat and recovery events test whether the same fixed graph can support
both re-differentiation and subsequent recoherence.

\subsection{Re-differentiation on a Fixed Graph}

For each retreat--recovery event, we use three checkpoints:
last-high, retreat, and recovery, defined below. Let $\Ablk(e)$ be the held-out block
accuracy at saved epoch $e$, and let $e_{\rm gen}$ be the generalization-onset
epoch of the same trajectory. We locate the retreat checkpoint first because the
other two are defined relative to it:
\begin{equation*}
 \begin{aligned}
 e_{\rm ret}&=\min\{e>e_{\rm gen}:\Ablk(e)<0.5\},\\
 e_{\rm high}&=\max\{e_{\rm gen}\leq e<e_{\rm ret}:\Ablk(e)\geq0.8\},\\
 e_{\rm rec}&=\min\{e>e_{\rm ret}:\Ablk(e)\geq0.8\},
 \end{aligned}
\end{equation*}
where $e_{\rm high}$, $e_{\rm ret}$, and $e_{\rm rec}$ are the last-high,
retreat, and recovery checkpoints, respectively. Recovery is absent if the
last set is empty. Table~\ref{tab:designs} summarizes this design. The training
data do not change over the 220 epochs of a training run, so neither does
their graph. It is connected and has $f_L=1$ at every checkpoint in the event. Because these checkpoints occur at different epochs across
seeds, we align trajectories by them rather than by absolute epoch. Across all 24
first-retreat events,
mean $\Kf$ increases from $1.273$ at last-high to $2.694$ at retreat, while mean $\wstar$ falls from
$0.880$ to $0.167$. For the 20 events with a saved recovery checkpoint, these
changes reverse: from retreat to recovery, mean $\Kf$ falls from $2.697$ to
$1.380$, while mean $\wstar$ rises from $0.182$ to $0.840$. 
These retreat means are over the 20 events where recovery occurs. 
Retreat is therefore a re-differentiation
of the learned function on fixed, complete identifiable support.

\subsection{Conditional Functions Split and Recombine}

The re-differentiation at retreat and the recoherence at recovery are changes
in how the 32 conditional functions group together. The increase in $\Kf$ 
shows that the Schmidt spectrum broadens, so the conditional functions 
become less alike. But the spectrum alone does not identify which left 
substrings separate from one another. We therefore cluster the
32 crossing-gauged conditional sign vectors $\{|\phi_a^\times\rangle\}_a$
defined in Eq.~\eqref{eq:gamma1-meaning} by similarity according 
to the details in Appendix~\ref{app:retreat-mechanism}, independently 
at last-high, retreat, and recovery when available. 
The clustering gives each of the 32 left substrings a cluster label at each
checkpoint. Let $T(e)$ be the label at saved epoch $e$ of a left substring drawn
uniformly from the 32. For
two checkpoints $e_0$ and $e_1$, we quantify cluster splitting and the change of the leading
subspace by
\begin{equation}
 \begin{aligned}
 B&=H[T(e_1)|T(e_0)]=-\sum_{r,s}\Pr[T(e_0)=r,T(e_1)=s]
  \log_2\frac{\Pr[T(e_0)=r,T(e_1)=s]}{\Pr[T(e_0)=r]},\\
 R_\chi&=\frac{1}{\chi}\Tr[\Pi_\chi(e_0)\Pi_\chi(e_1)].
 \end{aligned}
 \label{eq:retreat-structure}
\end{equation}
Here \(\Pr[T(e_0)=r,T(e_1)=s]\) is the fraction of the 32 left substrings 
assigned to cluster \(r\) at \(e_0\) and cluster \(s\) at \(e_1\). The marginal \(\Pr[T(e_0)=r]\) is defined analogously at \(e_0\).
$\Pi_\chi(e)$ is built from the eigenvectors of $\rho_A^\times(e)$. Ordering
its eigenvalues as $\lambda_1(e)\geq\lambda_2(e)\geq\cdots$ with eigenvectors
$|v_k(e)\rangle$, we set
\begin{equation*}
 \Pi_\chi(e)=\sum_{k=1}^{\chi}|v_k(e)\rangle\langle v_k(e)|,
 \qquad
 R_\chi=\frac{1}{\chi}\sum_{i,j=1}^{\chi}
 \bigl|\langle v_i(e_0)|v_j(e_1)\rangle\bigr|^2 .
\end{equation*}
The projector keeps the $\chi$ leading eigenvectors, 
so $R_\chi$ records how the leading subspace turns. 
The component projectors $\Pi_c$ of
Eq.~\eqref{eq:graph-dephasing} are built from basis vectors $|a\rangle$. $\Pi_\chi$ 
is instead built from eigenvectors. Thus the splitting entropy $B$ 
vanishes when no cluster splits and grows with the amount of splitting, 
while \(R_\chi\) equals one for identical subspaces and decreases as the subspaces 
rotate apart.

\begin{figure}[tbp]
  \centering
  \includegraphics[width=0.99\linewidth]{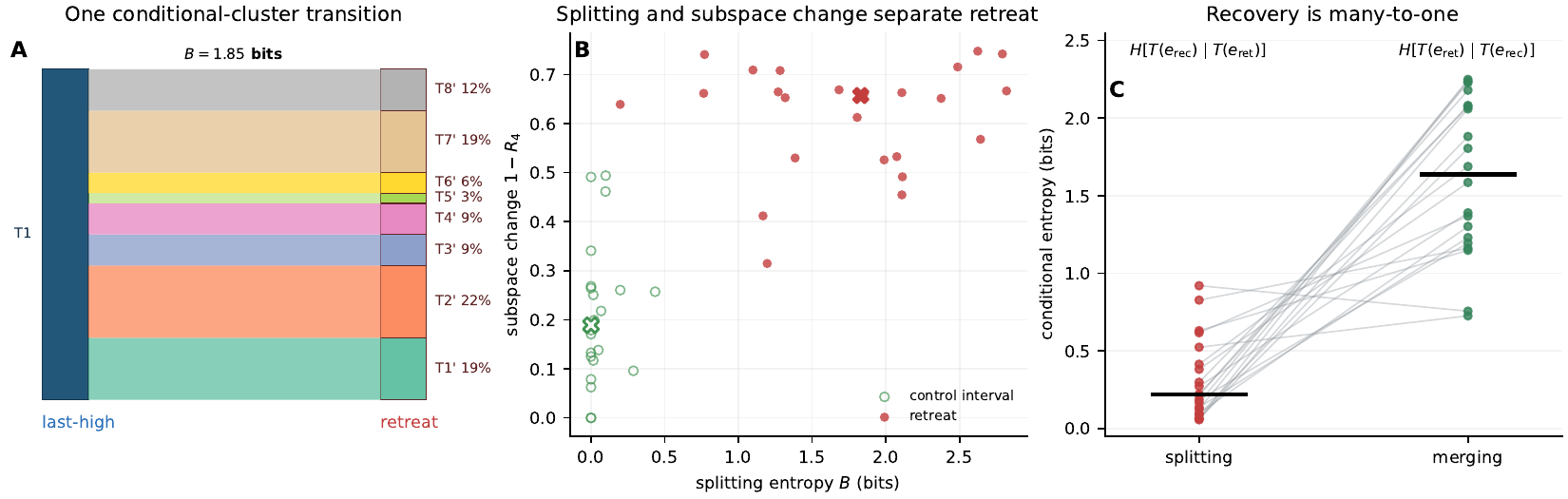}
  \caption{Change in conditional functions during retreat and recovery. (A)
  One representative retreat event. The clusters are those of output $j=8$, whereas $B$ is averaged over $j=8$--$11$. Each band runs from a last-high 
  cluster to a retreat
  cluster. Band heights are fractions of the 32 left substrings. 
  Cluster labels are defined
  independently at the two checkpoints. (B) Each point is one of the
  24 retreat events or its control interval, and crosses mark the
  median of each group. Retreat produces both more cluster splitting 
  (larger $B$) and a
  larger change of the leading four-dimensional subspace (larger $1-R_4$).
  (C) For each of the 20 recovered events, a line connects
  $H[T(e_{\rm rec})|T(e_{\rm ret})]$, which measures continued splitting, and
  $H[T(e_{\rm ret})|T(e_{\rm rec})]$, which measures many-to-one merging. Black bars mark
  medians.}
  \label{fig:lineage}
\end{figure}

We now present the results for $B$ and $R_\chi$ on the 24 first-retreat events.
In Fig.~\ref{fig:lineage}A, the 32 left substrings
occupy a single cluster at last-high and spread over eight clusters at retreat,
with $B=1.85$ bits averaged over $j=8$--$11$. For each retreat event, we select a nonoverlapping interval
from the same training run with the same duration, during which
held-out block accuracy remains above $\eta$. This interval
is the control interval in Fig.~\ref{fig:lineage}B.
Retreat has larger $B$ than 
its control interval in all 24 events. For
both $\chi=2$ and $4$, the last-high-to-retreat subspace overlap $R_\chi$ is
smaller than the overlap across the control interval in all 24 events.
Fig.~\ref{fig:lineage}B plots the case $\chi=4$.
Thus re-differentiation involves both splitting of previously
similar conditional functions and a change of their leading subspace.
Recovery reverses the cluster-level direction: conditional functions 
tend to recombine at recovery. Fig.~\ref{fig:lineage}C compares the two
directions.

\subsection{Recoherence without Exact Recurrence}
\label{sec:recurrence}

\begin{figure}[tbp]
  \centering
  \includegraphics[width=0.99\linewidth]{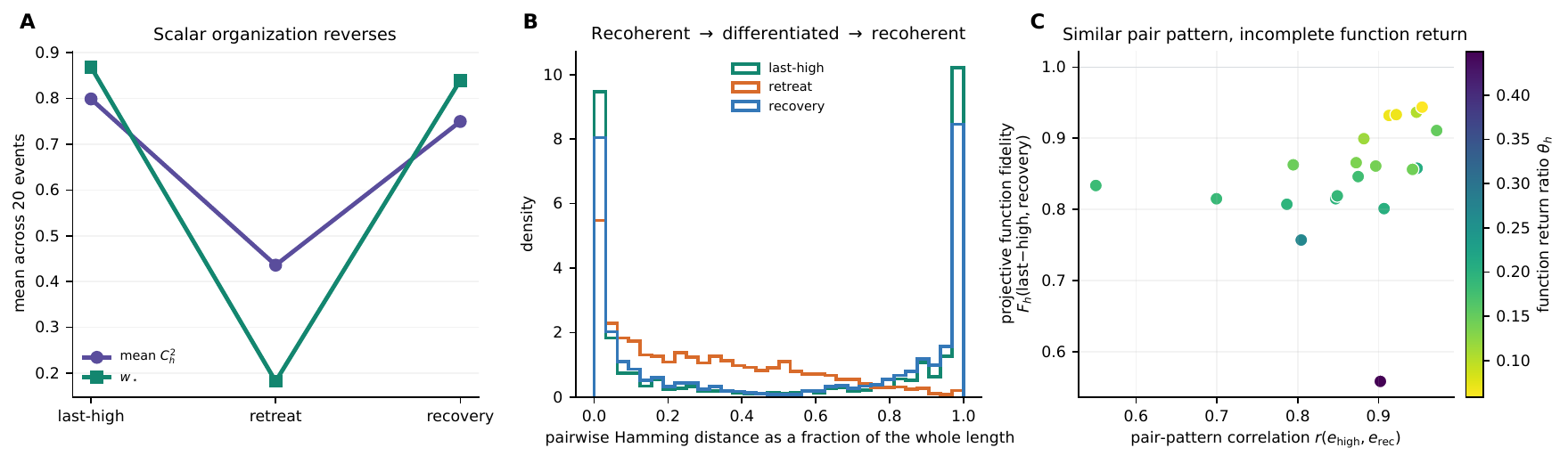}
  \caption{Retreat and recovery on the fixed connected graph for the 20
  events with a saved recovery. (A) Means of the off-diagonal $C_h^2$
  and $\wstar$ fall at retreat and return toward their last-high values at
  recovery. (B) For the conditional vector $\mathbf h_a=(h(a,b))_b$,
  the pairwise Hamming distances between $\mathbf h_a$ and $\mathbf h_{a'}$ 
  as fractions of the whole length move
  toward one half at retreat, indicating weaker row correlation, and return
  toward zero or one at recovery. The peaks at zero and one reflect the approximate
  factorization $h(a,b)\simeq u(a)v(b)$, for which
  $\mathbf h_{a'}\simeq\pm\mathbf h_a$.
  (C) Each point is one event averaged over the long-range outputs
  $j=8$--$11$. The horizontal coordinate is the pair-pattern
  correlation $r$ of Eq.~\eqref{eq:pair-pattern-correlation}. The vertical
  coordinate is $F_h(\text{last-high},\text{recovery})$, which stays below one and
  shows that the two sign tables differ. On each point, color shows the mean over the
  same long-range outputs of the residual function return ratio defined as
  $\theta_h$$=[1-F_h(\text{last-high},\text{recovery})]/
  [1-F_h(\text{last-high},\text{retreat})]$. Smaller values indicate a more complete return.
  One event sits apart at the lower right of the panel: its pair-pattern
  correlation is above the median of the 20, while its $F_h$ is the lowest and
  its $\theta_h$ the largest.}
  \label{fig:retreat-hierarchy}
\end{figure}

For the 20 events with a saved recovery checkpoint, the mean
$(\Kf,\wstar,\langle C_h^2\rangle_{a<a'})$ in the order
last-high $\to$ retreat $\to$ recovery is
\begin{equation}
 (1.300,0.869,0.799)\longrightarrow(2.697,0.182,0.436)
 \longrightarrow(1.380,0.840,0.750).
 \label{eq:retreat-fixed-events}
\end{equation}
Thus recovery restores a low-rank, target-aligned state.
Fig.~\ref{fig:retreat-hierarchy}A follows the means of
$\langle C_h^2\rangle_{a<a'}$ and $\wstar$ through the three checkpoints.

Fig.~\ref{fig:retreat-hierarchy}B resolves the same fall and return pair by
pair.
Write $\mathbf h_a=(h(a,b))_b$ for the conditional vector at left substring $a$.
For each pair, let $f$ be the Hamming distance between $\mathbf h_a$ and
$\mathbf h_{a'}$ as a fraction of the whole length. The coherence then follows as
$C_h(a,a')=1-2f$. Thus $f=1/2$ marks uncorrelated rows, while $f=0$ and $f=1$
mark $C_h(a,a')=\pm1$. These fractions spread
toward one half at retreat and concentrate again near zero and one at recovery.
The purity depends on $C_h^2$, so both ends indicate strong correlation.

To distinguish recovery of the pairwise correlations from recurrence of the full sign table at last-high, we compare the function-states at last-high and recovery using the
function fidelity
\begin{equation}
 F_h(t,t')=|\langle\psi_h(t)|\psi_h(t')\rangle|^2.
 \label{eq:function-fidelity}
\end{equation}
It equals one exactly when the two sign tables agree up to a global
sign.

This $F_h$ measures an angle between the function-states. Correspondingly, we define
an angle for the pair correlations alone. At saved epoch $e$ the pair
pattern $\mathbf p(e)$ collects the off-diagonal values $C_h(a,a')^2$ with
$a<a'$ into one vector. The fluctuation $\Delta\mathbf p(e)$ is the same vector
with its mean subtracted. The mean that is removed is the
scalar $\langle C_h^2\rangle_{a<a'}$ of Eq.~\eqref{eq:retreat-fixed-events},
which measures how much correlation is present. The fluctuation carries what
that scalar does not: which pairs are more correlated than average. We
compare
last-high with recovery by
\begin{equation}
 r(e_{\rm high},e_{\rm rec})=\frac{\Delta\mathbf p(e_{\rm high})\cdot\Delta\mathbf p(e_{\rm rec})}
 {|\Delta\mathbf p(e_{\rm high})|\,|\Delta\mathbf p(e_{\rm rec})|},
 \label{eq:pair-pattern-correlation}
\end{equation}
the cosine of the angle between the two fluctuations. It equals one when
the two patterns agree up to an overall scale and shift. 

Across the 20 recovered events, the last-high/recovery pair-pattern
correlations average $0.863$. Over the 80 recovered output states (four long-range outputs for each
of 20 events), the function fidelity square averages $0.846$ and remains below
one in every case. For each event, Fig.~\ref{fig:retreat-hierarchy}C plots the two angles
against each other, with $F_h$ on the vertical axis and $r$ on the horizontal. Recovery can therefore restore a similar low-rank,
target-selective organization while leaving residual differences in every
recovered sign table.

\FloatBarrier
\section{Discussion and Conclusion}
\label{sec:conclusion}

In this work, our results distinguish three aspects of learning: the constraint
graph fixed by data, the learned state, and predictive performance. Graph
connectivity specifies which relative signs between conditional functions are
identifiable from the training data.
Graph connectivity grows monotonically with
added distinct data size, whereas the function-state first differentiates and then
recoheres. That is, small additions of unseen data broaden the 
Schmidt spectrum, while
larger additions reconcentrate it toward the target mode. Branches that
differ only in whether the added data are unseen or duplicated show that this
reorganization follows the new constraints carried by unseen data rather than
continued training on data already seen. On the
already connected $+1{,}216$ graph, correlation magnitude $C^2_h$ changes first,
followed by target alignment, and the accuracy advantage emerges only
later. Thus the three aspects change at distinct stages of
learning. On a fixed graph, retreat and recovery show that the low-rank,
target-selective organization can be lost and restored without the same sign
table recurring.

The function-state construction allows the present task to be viewed as
quantum-state learning: at each checkpoint, the learned Boolean function
defines a pure equal-amplitude state, and training selects a state from this
family under the constraints imposed by the data. Purity, Schmidt spectra, target-mode population, and dephasing channels
are then exact properties of that state. 
Studies of neural representations of frustrated quantum ground states have found
that generalizing the signs can be substantially harder than generalizing the
amplitudes~\cite{westerhout2020generalization}. Boolean Fourier analysis provides
a complementary description of such sign structures
and relates their reconstruction complexity to learning
performance~\cite{schurov2025learning}.
More broadly, classical networks already
represent many-body states \cite{carleo2017solving} and reconstruct them from
measurements
\cite{torlai2018neural},
including single-shot images of Rydberg-atom tweezer arrays
\cite{torlai2019integrating}. Quantum
models learn states and operators directly
\cite{biamonte2017quantum}, given 
the operator--state correspondence, a
channel or an operator can be viewed as a state
\cite{cirac2001entangling}. Thus, in each of these settings
the learned state is constrained by data and it can be followed with the same purity, Schmidt spectrum, and dephasing
diagnostics. It is verifiable whether the learned state differentiates and then recoheres
as the data accumulate.

The present results concern a finite system and the parity function learned
by a Transformer. For larger systems, we expect the crossover from differentiation to
recoherence to shift with the string length as the connectivity thresholds of the
constraint graph do, which longer binary strings can test. Varying the architecture 
and the target function probes
whether the observed differentiation, recoherence, and predictive dynamics
persist in broader learning settings.
We expect the phenomenology to be broadly general. All these aspects are
defined for any learner whose answers to a fixed set of questions can be
recorded, whatever its architecture, its optimizer, and the function it 
is trained on. The same
ordering would then appear in any such learner as learning proceeds: 
the information needed 
is in place before the learner uses it,
the answers become more diversified before they simplify, 
and performance is the last
thing to improve. Along the way, a simple and sharp rule that
is lost would return through a different set of answers. 
Human learning can be described
this way, although how much of the mechanism is shared is an open question.

\section*{Code and data availability}

The code that trains the networks, performs the analyses and draws the figures
of this paper is available at \url{https://github.com/HanMaPhy/Learning_differentiation_and_recoherence}.
The trained networks and training records it reads are attached to the release
\texttt{data} of that repository.

\section*{Acknowledgments}

This work is funded
by the support from the ANR contract associated with the CNRS Chaire
Professeur Junior project ``QOQTIB''. H.M. also thanks École Polytechnique for
support.
I used ChatGPT and Claude Code to assist with coding, generating figures
and checking the grammar of the manuscript, and take full responsibility for the content of this paper.

\FloatBarrier
\appendix

\section{The Crossing Gauge}
\label{app:crossing-gauge}

Under the row and column flips $M(a,b)\mapsto e_af_bM(a,b)$ with
$e_a,f_b\in\{\pm1\}$, the reduced density matrix becomes $E\rho_AE$ with
$E=\operatorname{diag}_ae_a$. The population $w(u)$ of
Eq.~\eqref{eq:mode-population} along a mode chosen independently of $M$
therefore generally changes, whereas $\Kf$ is invariant. The uniform mode
$|u_{uni}\rangle$ is such a mode.

A small table with $n_L=n_R=4$, where $n_R=2^{D-d_L}$, shows the effect.
Flipping the second and third rows of $M$ gives $M'$,
\begin{equation*}
 M=\begin{pmatrix}+&+&-&-\\-&-&+&+\\-&-&+&+\\+&+&-&-\end{pmatrix},
 \qquad
 M'=\begin{pmatrix}+&+&-&-\\+&+&-&-\\+&+&-&-\\+&+&-&-\end{pmatrix}.
\end{equation*}
Both tables have rank one and $\Kf=1$. In $M'$ all rows are equal, so
$\langle u_{uni}|\rho_A|u_{uni}\rangle=1$. In $M$ the rows cancel in pairs, so
the same population is $0$. The two tables differ only by row signs, yet the
population along $|u_{uni}\rangle$ takes its two extreme values.

The crossing gauge of Eq.~\eqref{eq:crossing-gauge} removes this
dependence. In it, $0_A$ and $0_B$ are the all-zero left and right substrings 
and serve as the reference row and column. View the left and right substrings as the two vertex sets of a
complete bipartite graph, and $M(a,b)$ as the sign on the edge between $a$ and
$b$. The four
factors in Eq.~\eqref{eq:crossing-gauge} are the edges of the closed path
$a\to b\to0_A\to0_B\to a$. The product, and hence $M_\times$, is
unchanged by every flip. In the
example, every such product equals $+1$ because both tables have rank one. Thus
$M_\times=M'_\times$ is the all-ones table, and $\Gammone=1$ for both.


$\Gammone$ asks whether all crossing-gauged conditional functions share
one direction. When they form several groups, each group can have its own
direction. For a partition of the left substrings into $n$ clusters $c$ of
sizes $n_c$, we therefore define the cluster coherence
\begin{equation*}
 \Gamma(n)=\sum_c\frac{n_c}{n_L}\Bigl\|\frac{1}{n_c}\sum_{a\in c}|\phi_a^\times\rangle\Bigr\|^2 .
\end{equation*}
Each term is the squared length of the mean vector of one cluster, weighted by
the fraction of left substrings in that cluster. The cluster coherence equals
one when the vectors within every cluster coincide, and $\Gamma(1)=\Gammone$ by
Eq.~\eqref{eq:gamma1-meaning}. Appendix~\ref{app:retreat-mechanism} uses
$\Gamma(n)$ to choose the number of clusters.

\section{Protocol and Robustness}
\label{app:methods}

The Transformer is nominally overparameterized. We compare its $P=26{,}529$ trainable weights and biases
with the size of the training set. Per training input, $P/N=414.5$ at $N=64$
and $20.7$ at $N=1{,}280$. Each input carries one target bit at each of the
$D=11$ positions, and per target bit $P/(DN)=37.7$ and $1.88$. Even at
the largest dataset, $P$ therefore exceeds the number of target bits, 
which indicates overparameterization. These
ratios are parameter counts and do not estimate an effective model dimension.

During training we record the block accuracy of Eq.~\eqref{eq:ablk},
with the average over a fixed list of 4,096
inputs. Each entry is an 11-bit string with independent random bits, drawn
independently of the training set, so the list repeats strings and also
contains training inputs. The checkpoint-selection rules, such as the one
defining generalization onset, are applied to these recorded values.
All block accuracies also include the output at position $j=0$, whose
target bit is always zero.

The $N=64$ checkpoints are saved every 10 updates up to update 200, every
50 up to 1,000, and every 200 up to 4,000. The branching-study checkpoints are
at updates 0, 5, 10, 15, 20, 25, 50, 100, 200, 400, and 800 for the two
$+1{,}216$ branches and at updates 0, 25, 50, 100, 200, 400, and 800 for the
other branches. Bootstrap
intervals use 20,000 resamples. Each resample draws 25 starting models with
replacement from the original set of 25 and recomputes the mean. 
The interval spans the 2.5th
and 97.5th percentiles of the 20,000 resampled means.

\subsection{Added-Data Robustness}
\label{app:causal-support}

The distinct additions follow one random ordering of the 1,984 inputs
absent from the $N=64$ training set, nested as in the caption of
Fig.~\ref{fig:constraint}.
To check that the result does not depend on this ordering, we draw two further
independent orderings and repeat the $+192$, $+576$ and $+1{,}216$ branches for eight
of the starting models. Together with the ordering of Sec.~\ref{sec:causal}, this gives 24
combinations of starting model and ordering at each size. In all of them at $+576$ and $+1{,}216$ the
distinct-minus-duplicate contrast $\Delta\Kf$ is negative at update 800, so the
large-addition recoherence does not depend on the particular ordering.
At $+192$ it is negative in 18 of the 24.

\subsection{Clustering of Conditional Functions}
\label{app:retreat-mechanism}

The splitting entropy $B$ of Eq.~\eqref{eq:retreat-structure} requires a
cluster label for each of the 32 left substrings at each checkpoint. We obtain
the labels by grouping the 32 crossing-gauged conditional functions
$|\phi_a^\times\rangle$ of Eq.~\eqref{eq:gamma1-meaning}. Each is a vector of
$2^{D-d_L}=64$ entries with unit length, and we group them by Euclidean
distance, whose square is
$\||\phi_a^\times\rangle-|\phi_{a'}^\times\rangle\|^2=2-2C_h^\times(a,a')$ with
$C_h^\times$ of Eq.~\eqref{eq:crossing-gram}. Since the entries are $\pm1$,
this squared distance is $4d_H/2^{D-d_L}$, where $d_H$ is the Hamming distance
between rows $a$ and $a'$ of $M_\times$. The same ratio $d_H/2^{D-d_L}$,
computed without the gauge, is plotted in Fig.~\ref{fig:retreat-hierarchy}B.
The crossing gauge of Appendix~\ref{app:crossing-gauge} is needed because a row
flip sends $|\phi_a\rangle$ to $-|\phi_a\rangle$ while leaving the Schmidt
spectrum, $\Kf$, and $C_h^2$ unchanged. A grouping computed from the ungauged
$|\phi_a\rangle$ would therefore depend on signs that these quantities do not
resolve.

For a given number of clusters $n$, we use the method $k$-means with
$k=n$. A center is a point of the same $64$-dimensional space and is not one of the
$|\phi_a^\times\rangle$. Starting from $n$ centers, the
algorithm assigns each vector to its nearest center. It then moves each center
to the mean of the vectors assigned to it. The two steps repeat until the
centers stop moving, for at most 80 iterations. The outcome depends on the starting centers. We choose them by
$k$-means++. The first center is a randomly chosen vector, and each further center is
drawn with probability proportional to its squared distance
from the nearest center already chosen. We run the procedure from 12
independent starts and keep the partition with the smallest summed squared
distance between the vectors and their centers. This sum equals
$n_L[1-\Gamma(n)]$ with the cluster coherence $\Gamma(n)$ of
Appendix~\ref{app:crossing-gauge}, so the partition we keep is the one with the
largest $\Gamma(n)$.

We increase $n$ from 1 to 8 and keep the first partition whose cluster
coherence $\Gamma(n)$ reaches $0.90$. If none does, we keep the partition with
$n=8$. Clustering is carried out independently at each checkpoint and for each
of the four long-range outputs $j=8$--$11$. The quantities $B$ and $R_\chi$ are
computed for each output and then averaged over the four.

\begin{figure}[tbp]
  \centering
  \includegraphics[width=0.99\linewidth]{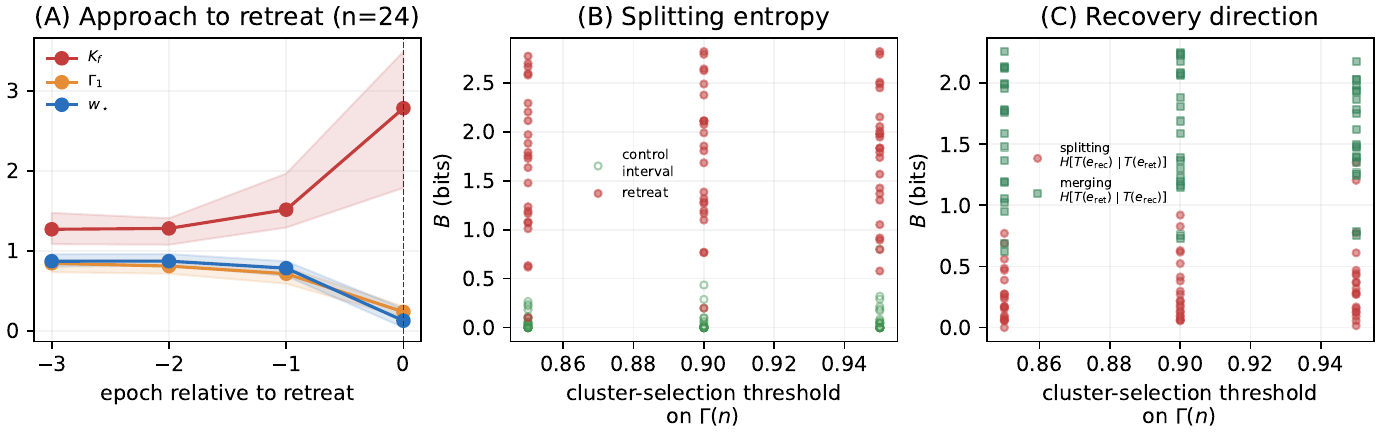}
  \caption{Robustness of the retreat--recovery result. (A) Median $\Kf$,
  $\Gammone$, and $\wstar$ with interquartile shading for all 24 first-retreat
  events over the four epochs ending at retreat. (B) Splitting entropy at retreat,
  $B=H[T(e_{\rm ret})|T(e_{\rm high})]$, and over the control interval,
  $H[T(e_1)|T(e_0)]$ with $e_0$ and $e_1$ its endpoints, at thresholds 0.85,
  0.90, and 0.95 on $\Gamma(n)$. (C) The same quantity for the
  retreat--recovery pair in both orders: merging
  $H[T(e_{\rm ret})|T(e_{\rm rec})]$ exceeds splitting
  $H[T(e_{\rm rec})|T(e_{\rm ret})]$ in 19 of the 20 events at every threshold.}
  \label{fig:robustness}
\end{figure}

Fig.~\ref{fig:robustness} adds two views of the retreat analysis. Panel A
resolves $\Kf$, $\Gammone$, and $\wstar$ epoch by epoch over the four epochs
ending at retreat, instead of at the last-high and retreat checkpoints alone.
All three change little over the first three of those epochs and then move
sharply at the last one, which carries $84$, $79$, and $88$ per cent of the
total median change in $\Kf$, $\Gammone$, and $\wstar$. Panels B and C repeat
the clustering with the threshold on $\Gamma(n)$ set to $0.85$, $0.90$, and
$0.95$. Of these, $0.90$ is the value used everywhere else.

\begingroup
\small
\bibliographystyle{unsrt}
\bibliography{references}
\endgroup

\end{document}